\documentclass[aps,floats]{revtex4}
\usepackage{amsmath,amssymb}
\usepackage{graphicx,epsfig}
\begin{document}
\bibliographystyle {plain}

\def\oppropto{\mathop{\propto}} 
\def\opsimeq{\mathop{\simeq}}
\def\opoverderline{\mathop{\overline}}
\def\operarrow{\mathop{\longrightarrow}}
\def\opsim{\mathop{\sim}}

\def\fig#1#2{\includegraphics[height=#1]{#2}}
\def\figx#1#2{\includegraphics[width=#1]{#2}}

%\newcommand{\fig}[2]{\epsfxsize=#1\epsfbox{#2}} \reversemarginpar 

%%%%%%%%%%%%%%%%%%%%%%%%%%%%%%%%%%%%%%%%%%%%%%%%%%%%%%%%%%%%%%%%%%%%%%%%%%%%
\title{ Strong Disorder RG principles within a fixed cell-size real space renormalization  : \\ 
application to the Random Transverse Field Ising model on various fractal lattices  } 

%%%%%%%%%%%%%%%%%%%%%%%%%%%%%%%%%%%%%%%%%%%%%%%%%%%%%%%%%%%%%%%%%%%%%%%%%%%%

 \author{ C\'ecile Monthus and Thomas Garel }
  \affiliation{Institut de Physique Th\'{e}orique, 
CNRS and CEA Saclay, 
 91191 Gif-sur-Yvette cedex, France}

\begin{abstract}
Strong Disorder Renormalization is an energy-based renormalization that leads to a complicated renormalized topology for the surviving clusters as soon as $d>1$. In this paper, we propose to include Strong Disorder Renormalization ideas within the more traditional fixed cell-size real space RG framework. We first consider the one-dimensional chain as a test for this fixed cell-size procedure: we find that all exactly known critical exponents are reproduced correctly, except for the magnetic exponent $\beta$ (because it is related to more subtle persistence properties of the full RG flow). We then apply numerically this fixed cell-size procedure to two types of renormalizable fractal lattices (i) the Sierpinski gasket of fractal dimension $D=\ln 3/\ln 2$, where there is no underlying classical ferromagnetic transition, so that the RG flow in the ordered phase is similar to what happens in $d=1$ (ii) a hierarchical diamond lattice of fractal dimension $D=4/3$, where there is an underlying classical ferromagnetic transition, so that the RG flow in the ordered phase is similar to what happens on hypercubic lattices of dimension $d>1$. In both cases, we find that the transition is governed by an Infinite Disorder Fixed Point : besides the measure of the activated exponent $\psi$, we analyze the RG flow of various observables in the disordered and ordered phases, in order to extract the 'typical' correlation length exponents of these two phases which are different from the finite-size correlation length exponent.

\end{abstract}

\maketitle

\section{ Introduction }

The choice to work in real-space to define renormalization procedures, 
which already presents a great interest for pure systems
 \cite{realspaceRG},
becomes the unique choice for 
 disordered systems if one wishes to describe spatial heterogeneities.
Whenever these disorder heterogeneities play a dominant role 
over thermal or quantum fluctuations, the most appropriate renormalization
procedures
are Strong Disorder renormalizations \cite{StrongRGreview}
that have been introduced by Ma and Dasgupta \cite{Ma-Dasgupta} : as shown by
Daniel Fisher \cite{danieltransverse,danielAF}, these Strong Disorder RG rules
lead to asymptotic exact results if the broadness of the disorder
distribution grows indefinitely at large scales.
In dimension $d=1$, exact results for a large number of observables
have been explicitly computed \cite{danieltransverse,danielAF} because the renormalized lattice of surviving
degrees of freedom remains one-dimensional.
In dimension $d>1$, the Strong Disorder RG procedure can still be defined,
but it cannot be solved analytically, 
because the topology of the lattice changes upon renormalization.
Nevertheless, Strong Disorder RG rules can be implemented numerically.
For instance, for the disordered Quantum Ising model,
these numerical RG studies have concluded that the transition is also governed by
an Infinite Disorder fixed point in dimensions $d=2,3,4$ 
 \cite{fisherreview,motrunich,lin,karevski,lin07,yu,kovacsstrip,
kovacs2d,kovacs3d,kovacsentropy,kovacsreview}, 
 in agreement with the results of independent quantum Monte-Carlo
in $d=2$ \cite{pich,rieger}.

Nevertheless, the complicated topology that emerges between renormalized degrees of freedom
in dimension $d>1$ tends to obscure the physics, because a large number of very weak bonds are generated
during the RG, that will eventually not be important for the forthcoming RG steps.
In this article, we propose to include strong disorder RG ideas
within the more traditional fixed-length-scale real space RG framework
that preserves the topology upon renormalization. In particular for renormalizable fractal lattices
like the Sierpinski gasket or diamond hierarchical lattices, this fixed-length-scale RG procedure
allows to use the so-called 'pool method' and to study numerically very large system sizes.
(Note that when the full strong disorder renormalization is applied to fractal lattices
like the Sierpinski gasket as in Ref \cite{melin}, 
the system sizes are limited because one cannot take advantage of the self-similarity
of the original lattice which is immediately broken by the full RG.)

The paper is organized as follows.
In section \ref{sec_fixed}, we introduce the principles of 
Strong Disorder RG within a fixed cell-size RG framework.
In section \ref{sec_nume1d}, we present our numerical result for $d=1$, and conclude
that this procedure correctly captures all critical exponents except for the 
magnetic exponent $\beta$ which is related to persistence properties of the full RG flow.
In section \ref{sec_triangle}, we apply numerically our procedure to the Sierpinski gasket
of fractal dimension $D=\ln 3/\ln 2=1.58..$ with no underlying classical ferromagnetic transition.
In section \ref{sec_b2c8}, we study numerically the case of a diamond hierarchical lattice
presenting an underlying classical ferromagnetic transition.
Section \ref{sec_conclusion} summarizes our conclusions.
Appendix \ref{sec_full} contains a reminder on the usual Strong Disorder RG on arbitrary lattices
and on the properties of renormalized observables as a function of the energy-based RG scale $\Gamma$.

\section{ Strong Disorder RG within a fixed cell-size real-space RG framework}

\label{sec_fixed}

In this paper, we consider the quantum Ising model defined in terms of Pauli matrices
\begin{eqnarray}
{\cal H} =  -  \sum_{<i,j>} J_{i,j}  \sigma^z_i \sigma^z_j - \sum_i h_i \sigma^x_i
\label{hdes}
\end{eqnarray}
where the nearest-neighbor couplings $J_{i,j}>0$
and the transverse fields $h_i>0$ are independent random variables
drawn with two distributions $\pi_{coupling}(J)$ and $\pi_{field}(h)$.

As recalled in Appendix \ref{sec_full}, the Strong Disorder Renormalization
for the quantum Ising model of Eq. \ref{hdes}
 is {\it an energy-based RG} , 
where the strongest ferromagnetic bond or the strongest transverse field
is iteratively eliminated. In dimension $d>1$ this leads to a complicated topology
for the network of surviving clusters at large RG scale $\Gamma$.
In this section, we propose to apply the Strong Disorder RG principles 
{\it within a fixed cell-size framework }, i.e. the RG scale will not be an energy-based scale like $\Gamma$,
but a length-scale $L$ as in usual real-space RG procedures.
Let us first describe the expected scalings as a function of $L$ in various phases,
to justify the possibility of such a fixed-length procedure.

\subsection{ Expected scaling of renormalized observables as a function of the size $L$ }

Since the energy-RG-scale $\Gamma$ is associated to some length-scale $l_{\Gamma}$ via Eq. \ref{ngamma},
it seems clear that all RG flows as a function of $\Gamma$ described in Appendix \ref{sec_full}
 can be reformulated as RG flows as a function of the length scale $L$.
Physically, these scalings in $L$ simply describe the scaling of observables of finite-size samples.
We should stress however that the change from the energy-scale $\Gamma$ to the length-scale $L$ is not
 just a 'mathematical change of variables',
because probability distributions are involved. Indeed, the ensemble of disordered samples of volume $L^d$
is characterized by a probability distribution $P_L(\Gamma_{last})$ of the last RG scale $\Gamma_{last}$.
Similarly for a sample in the thermodynamic limit $L \to +\infty$ in $d=1$, the state at RG scale $\Gamma$
is characterized by probability distribution of
 lengths $(l_B,l_C)$ for renormalized bonds and renormalized clusters \cite{danieltransverse}
(in dimension $d>1$, the state at RG scale $\Gamma$ involves in addition
a complicated topology of the renormalized surviving clusters).

\subsubsection{ Critical Point  : RG flow as a function of the size $L$  }

At the Infinite Disorder critical point described in section \ref{sec_criti},
 the renormalized transverse fields and the renormalized
couplings remain in competition at all scales,
i.e. they display the same activated scaling in $L$ with some exponent $\psi$
\begin{eqnarray}
\ln J_L && = -  L^{\psi} u_c \nonumber \\
\ln h_L && = -  L^{\psi} v_c
\label{JLhLcriti}
\end{eqnarray}
where $(u_c,v_c)$ are $O(1)$ random variables.
The exponent $\psi$ corresponds to 
the activated exponent of the gap $G(L) \propto e^{-L^{\psi}}$ (Eq \ref{defpsicriti}).

\subsubsection{ Critical Region  : Finite-Size Scaling with the exponent $\nu_{FS}$  }

At this Infinite Disorder fixed point, one expects
the following finite-size scaling form for the typical values \cite{motrunich}
\begin{eqnarray}
\ln J_L^{typ} \equiv \overline{ \ln J_L }  = -  L^{\psi} F_J\left( L^{1/\nu_{FS}} \vert h-h_c \vert \right) \nonumber \\
\ln h_L^{typ} \equiv \overline{ \ln h_L }  = -  L^{\psi} F_h\left( L^{1/\nu_{FS}} \vert h-h_c \vert \right)
\label{fsslnjl}
\end{eqnarray}
as well as for the width of the distribution of $(\ln J_L)$ 
\begin{eqnarray}
\Delta_L  \equiv  \left( \overline{ (\ln J_L)^2 } - (\overline{ \ln J_L })^2 \right)^{1/2} 
= L^{\psi} F_{\Delta} \left( L^{1/\nu_{FS}} \vert h-h_c \vert \right)
\label{defdeltaL}
\end{eqnarray}
(and similarly for the width of distribution of $(\ln h_L)$).
Here $\nu_{FS}$ is introduced as the correlation length exponent that govern all finite-size effects in
the critical region.

The magnetization $\mu_L$ of critical clusters grows
 with the fractal dimension $d_f$ introduced in Eq \ref{mucriti}. The intensive magnetization 
is expected to follow the usual power-law finite-size scaling form
\begin{eqnarray}
m  \equiv \frac{\mu_L}{ L^d } \propto L^{-x} F_M \left( L^{1/\nu_{FS}} \vert h-h_c \vert \right)
\ \  {\rm with} \ \  x=d-d_f
\label{mcriti}
\end{eqnarray}

\subsubsection{ Disordered Phase : RG flow as a function of the size $L$  }

In the disordered phase described in section \ref{sec_disorder}, the transverse fields $h_i$
are not renormalized anymore asymptotically, i.e. they converge towards finite values as $L \to +\infty$
\begin{eqnarray}
\ln h_{\infty}^{typ} \equiv \overline{\ln h_{\infty}} \propto   - (h-h_c)^{-\kappa} 
\label{defkappa}
\end{eqnarray}
where the exponent $\kappa$ of the essential singularity satisfies
\begin{eqnarray}
\kappa = \psi \nu_{FS}
\label{kappafss}
\end{eqnarray}
as a consequence of the matching with the finite-size scaling form of Eq. \ref{fsslnjl}.

The renormalized couplings $J_{L}$ is expected to have the same scaling as the two-point correlation function
\cite{danieltransverse,motrunich}
\begin{eqnarray}
\ln J_L = - \frac{L}{\xi_{typ}} + L^{\omega} A(h) u
\label{JLdisorder}
\end{eqnarray}
The first term is non-random and describes the exponential decay with the size $L$,
where $\xi_{typ} $ represents the typical correlation length
\begin{eqnarray}
\ln J_L^{typ} \equiv \overline{\ln J_L} \oppropto_{L \to +\infty} - \frac{L}{\xi_{typ}} 
\label{JLdisordertyp}
\end{eqnarray}
  The compatibility 
with the finite-size scaling form of Eq. \ref{fsslnjl} implies that the typical correlation exponent
$\nu_{typ}$ is different from the finite-size correlation length exponent $\nu_{FS}$ and reads
\begin{eqnarray}
\nu_{typ} = (1-\psi) \nu_{FS}
\label{nutyp}
\end{eqnarray}
The exponent $\nu_{FS}$ is expected \cite{danieltransverse,motrunich}
to correspond to the exponent $\nu_{av}$
of the {\it averaged } two-point correlation function ($\nu_{FS}=\nu_{av}$).

The second term in Eq \ref{JLdisorder} contains an $O(1)$ random variable $u$.
It is usually subleading with respect to the first term,
i.e. of order $L^{\omega}$ with some exponent $\omega<1$.
We have argued in \cite{us_transverseDP} that
this exponent $\omega$ should coincide with the droplet exponent $\omega_{DP}(D=d-1)$
 of the Directed Polymer with $D=(d-1)$ transverse directions.
The compatibility with the finite-size scaling form of Eq. \ref{defdeltaL}
for the width $\Delta_L$ of the distribution of $\ln J_L$ 
\begin{eqnarray}
\Delta_L  \equiv  \left( \overline{ (\ln J_L)^2 } - (\overline{ \ln J_L })^2 \right)^{1/2} 
\oppropto_{L \to \infty}   A(h) L^{\omega}
\label{deltaLdes}
\end{eqnarray}
implies that the amplitude $A(h)$ is non-singular as $h \to h_c$ only if $\psi=\omega$,
which is known to be the case in $d=1$ \cite{danieltransverse}.
On the other hand, if the two exponents turn out to be different $\psi \ne \omega$,
then the amplitude $A(h)$ has to present the following power-law singularity 
\begin{eqnarray}
A(h) \oppropto_{h \to h_c^+} (h-h_c)^{- (\psi-\omega) \nu_{FS}}
\label{defampliA}
\end{eqnarray}

\subsubsection{ Ordered Phase : RG flow as a function of the size $L$  }

In the ordered phase described in section \ref{sec_order}, 
the magnetization $\mu$ of surviving clusters grows extensively
\begin{eqnarray}
\mu_L \propto   L^d 
\label{muLorder}
\end{eqnarray}
i.e. the intensive magnetization is finite and vanishes with some power-law singularity 
compatible with the finite-size scaling form of Eq. \ref{mcriti}
\begin{eqnarray}
m \equiv \frac{\mu_L}{ L^d } \propto (h_c-h)^{\beta} \ \ \ {\rm with } \ \ \beta=x \nu_{FS}
\label{defbeta}
\end{eqnarray}

Since surviving clusters have an extensive magnetization, 
the logarithm of their renormalized transverse fields grows also extensively
\begin{eqnarray}
\ln h_L \oppropto_{L \to \infty}   - \left( \frac{L}{\xi_h} \right)^d v
\label{hLorder}
\end{eqnarray}
where $v$ is an $O(1)$ random variable.
The length scale $\xi_h$ represents the characteristic size of finite disordered clusters
within this ordered phase. The compatibility
with the finite-size scaling form of Eq. \ref{fsslnjl} implies that the typical correlation exponent
$\nu_{h}$ reads
\begin{eqnarray}
\nu_{h} = \left( 1- \frac{\psi}{d}\right) \nu_{FS}
\label{nuh}
\end{eqnarray}
Note that $\nu_h$ plays in the ordered phase a role similar to $\nu_{typ}$ in the disordered phase, but that they coincide only if $d=1$ (Eq \ref{nutyp}).

For the asymptotic behavior of the renormalized couplings $J_{L}$ between surviving clusters,
one needs first to determine whether there exists an underlying classical ferromagnetic transition, as recalled in section \ref{sec_order}. 

(a) When there is no underlying classical ferromagnetic transition, the typical renormalized coupling remains finite, and presents the same essential singularity as in Eq. \ref{defkappa}
\begin{eqnarray}
\ln J_{\infty}^{typ} \equiv \overline{\ln J_{\infty}} \propto    - (h_c-h)^{-\kappa} 
\label{noclassical}
\end{eqnarray}

(b) When there exists
 an underlying classical ferromagnetic transition,
the renormalized couplings $J_L$ do not remain finite as in Eq. \ref{noclassical}
 but grow asymptotically at large $L$
with the scaling of the {\it classical } random ferromagnet model
\begin{eqnarray}
J_L \oppropto_{L \to +\infty}  \sigma L^{d_s} +  L^{\theta_s} u
\label{jijclassical}
\end{eqnarray}
The first non-random term grows as the surface $L^{d_s}$ of dimension $d_s$
($\sigma$ being the surface tension). The second term contains a random variable $u$
and grows with some exponent $\theta_s$.
For instance in $d=2$ where the surface of dimension $d_s=d-1=1$ is actually a line,
the exponent $\theta_s$ coincides with the droplet exponent $\omega_{DP}=1/3$
of the Directed Polymer in dimension $1+1$.
In dimension $d=3$ where the surface has dimension $d_s=d-1=2$, the exponent $\theta_s$
has been numerically measured to be $\theta_s \simeq 0.84$ \cite{middleton}.
It is clear that in the regime of large $L$ where Eq. \ref{jijclassical} holds,
the strong disorder RG procedure is not appropriate anymore,
because the couplings $J$ grow with the scale $L$ (so the decimation of the biggest coupling
tend to create a bigger renormalized coupling), and because the width of the probability
distribution of $(\ln J_L)$ actually decrease with $L$ (instead of becoming broader and broader)
\begin{eqnarray}
\Delta_L   \equiv  \left( \overline{ (\ln J_L)^2 } - (\overline{ \ln J_L })^2 \right)^{1/2} 
\oppropto_{L \to \infty}  L^{-d_s+\theta_s} 
\label{deltajijclassical}
\end{eqnarray}
This is why in the language of the energy scale $\Gamma$, the strong disorder RG
stops at some finite value $\Gamma_{perco}$ where percolation occurs.
In the language of the length scale $L$, the RG flow exists for all $L$
but is non-monotonous in the critical region of the ordered phase :
at the beginning, the typical coupling decays in Eq. \ref{JLhLcriti}
and $\Delta_L$ grows as $L^{\psi}$, whereas asymptotically the typical coupling
grows as in Eq. \ref{jijclassical} and $\Delta_L$ decays as in Eq. \ref{deltajijclassical}.

\subsubsection{ Discussion  }

The fact that RG flows can be reformulated in terms of the length-scale $L$
suggests that some fixed cell-size real-space RG procedure should be possible
even for Infinite Disorder fixed points.
In the following, we propose such an explicit RG procedure, where Strong Disorder decimation rules
are used within a fixed cell-size framework.

\subsection{ Principles  of the fixed cell-size RG procedure for the one-dimensional chain }

\label{sec_principles}

For clarity, we first explain the ideas on the case of the one-dimensional chain,
and compare with the exact results of the energy-scale strong disorder RG \cite{danieltransverse}.

\subsubsection{ Notations }

\label{sec_notations}

Let us first introduce notations.
At generation $n$, corresponding to a length $L_n=b^n$,
a {\it renormalized open bond} $]A,B[$ is characterized by
the correlated renormalized variables $(J_{AB}, r_A,\Delta \mu_A ; r_B,\Delta \mu_B)$ 
distributed with some joint probability distribution
$P_n( J_{AB}, r_A,\Delta \mu_A ; r_B,\Delta \mu_B)$ :

(i) $J_{AB}$ represents the renormalized ferromagnetic coupling between the two end points $A$ and $B$

(ii) $r_A$ represents the multiplicative 
factor that comes from the interior of the bond 
and that should be applied to the transverse field of $A$, see the RG rule of Eq. A3.
(Similarly $r_B$ concerns the other end $B$ of the bond).

(iii) $\Delta \mu_A$ represents the excess of magnetization 
that comes from the interior of the bond and that should be added
to the magnetization of $A$, see the RG rule of Eq. A4. 
(Similarly $\Delta \mu_B$ concerns the other end $B$ of the bond).

At generation $n=0$ : the coupling  $J_{AB}$ is just an initial coupling of the model
drawn with some disorder distribution $ \pi_{coupling}(J_{AB})$;
the sites $A$ and $B$ have magnetic moments $\mu_A=1=\mu_B$, 
corresponding to $\Delta \mu_A=0=\Delta \mu_B$ ; and the transverse fields $h_A^{(0)}$ 
and $h_B^{(0)}$ are the initial transverse fields of the model,
corresponding to $r_A=1=r_B$.
So the initial joint-distribution is simply
\begin{eqnarray}
P_{n=0}( J_{AB}, r_A,\Delta \mu_A ; r_B,\Delta \mu_B)
= \pi_{coupling}(J_{AB}) \delta(r_A-1) \delta(\Delta \mu_A) 
\delta(r_B-1) \delta(\Delta \mu_B)
\label{p0initial}
\end{eqnarray}

\subsubsection{ Step 1 }

\label{sec_step1}

The first step consists in building the elementary structure of generation $(n+1)$
from the renormalized open bonds of generation $n$.

For the one-dimensional chain with a scaling factor $b$, the explicit construction is as follows :

1a) Draw $b$ independent renormalized open bonds at generation $n$,
where each open bond $j=1,2,..,b$ is characterized by the variables
$(J_{A_jB_j}, r_{A_j},\Delta \mu_{A_j} ; r_{B_j},\Delta \mu_{B_j}$.

1b) For each $j=1,2,..(b-1)$, connect the two open bonds $(j)$ and $(j+1)$
in series via the introduction of an intermediate site $C_j$
corresponding to the merging of $B_j$ with $A_{j+1}$.
The intermediate site $C_j$ has thus for magnetization 
the sum of its own initial magnetization $\mu_{C_j}^{(0)}=1 $
and of the excesses $\Delta \mu_{A_{j+1}}$ and $\Delta \mu_{B_j} $  coming from the renormalization
of the interiors
of the two neighboring bonds
\begin{eqnarray}
\mu_{C_j}=\mu_{C_j}^{(0)}+ \Delta \mu_{A_{j+1}} + \Delta \mu_{B_j}
\label{muCjd1}
\end{eqnarray}
Similarly its transverse field is the product of its own initial transverse field $h_{C_j}^{(0)}$
and of the two possible reducing factors $r_{A_{j+1}}$ and $ r_{B_j}$ coming from
the renormalization
of the interiors
of the two neighboring bonds
\begin{eqnarray}
h_{C_j}=h_{C_j}^{(0)}\ r_{A_{j+1}} \ r_{B_j}
\label{hCjd1}
\end{eqnarray}

The end-point $A$ of the new bond of the $(n+1)$ generation
will correspond to the end-point $A_1$ of the bond $j=1$,
and inherits the associated variables 
\begin{eqnarray}
\Delta \mu_{A}^{(step \ 1)}= \Delta \mu_{A_1}
\label{muastep1}
\end{eqnarray}
and 
\begin{eqnarray}
r_{A}^{(step \ 1)}= r_{A_1} 
\label{rastep1}
\end{eqnarray}
Similarly,  the end-point $B$ of the new bond of the $(n+1)$ generation
will correspond to the end-point $B_b$ of the bond $j=b$
and inherits the associated variables 
\begin{eqnarray}
\Delta \mu_{B}^{(step \ 1)}= \Delta \mu_{B_b}
\label{mubstep1}
\end{eqnarray}
and 
\begin{eqnarray}
r_{B}^{(step \ 1)}= r_{B_b} 
\label{rbstep1}
\end{eqnarray}

\subsubsection{ Step 2 }

\label{sec_step2}

The second step consists in applying the usual strong disorder RG rules
 (recalled in section \ref{SDRGfullrules}) 
to the internal structure with renormalized bonds of generation $n$
that has been constructed in step 1, up to the final state containing
a single renormalized open bond of generation $(n+1)$.

For the one-dimensional chain with a scaling factor $b$, the explicit strong disorder
renormalization of the internal structure is as follows :

These two end-points $A$ and $B$ of the new open bond of the $(n+1)$ generation are non-decimable by definition,
whereas all $(b-1)$ intermediate points $(C_1,C_2,..,C_{b-1})$, and all internal bonds $(J_{AC_1}=J_{A_1B_1},
J_{C_1C_2}=J_{A_2B_2},...,J_{C_{b-1}B}=J_{A_bB_b}$ are decimable.

We apply to this internal structure the usual strong disorder RG rules recalled in Appendix A1.
So we look for the maximum between the $b$ couplings and the $(b-1)$ transverse fields
\begin{eqnarray}
\Omega= {\rm max } \left[J_{AC_1},h_{C_1},J_{C_1C_2},h_{C_2},..,h_{C_{b-1}},J_{C_{b-1}B} \right]
\label{omegab}
\end{eqnarray}
and we apply iteratively the strong disorder RG rules up to the full decimation
of this internal structure : the final output is the the joint variables
$(J_{AB}, r_A,\Delta \mu_A ; r_B,\Delta \mu_B)$ of an open bond of generation $(n+1)$.

\subsubsection{ Explicit RG for the rescaling factor $b=2$ }

\begin{figure}[htbp]
\includegraphics[height=8cm]{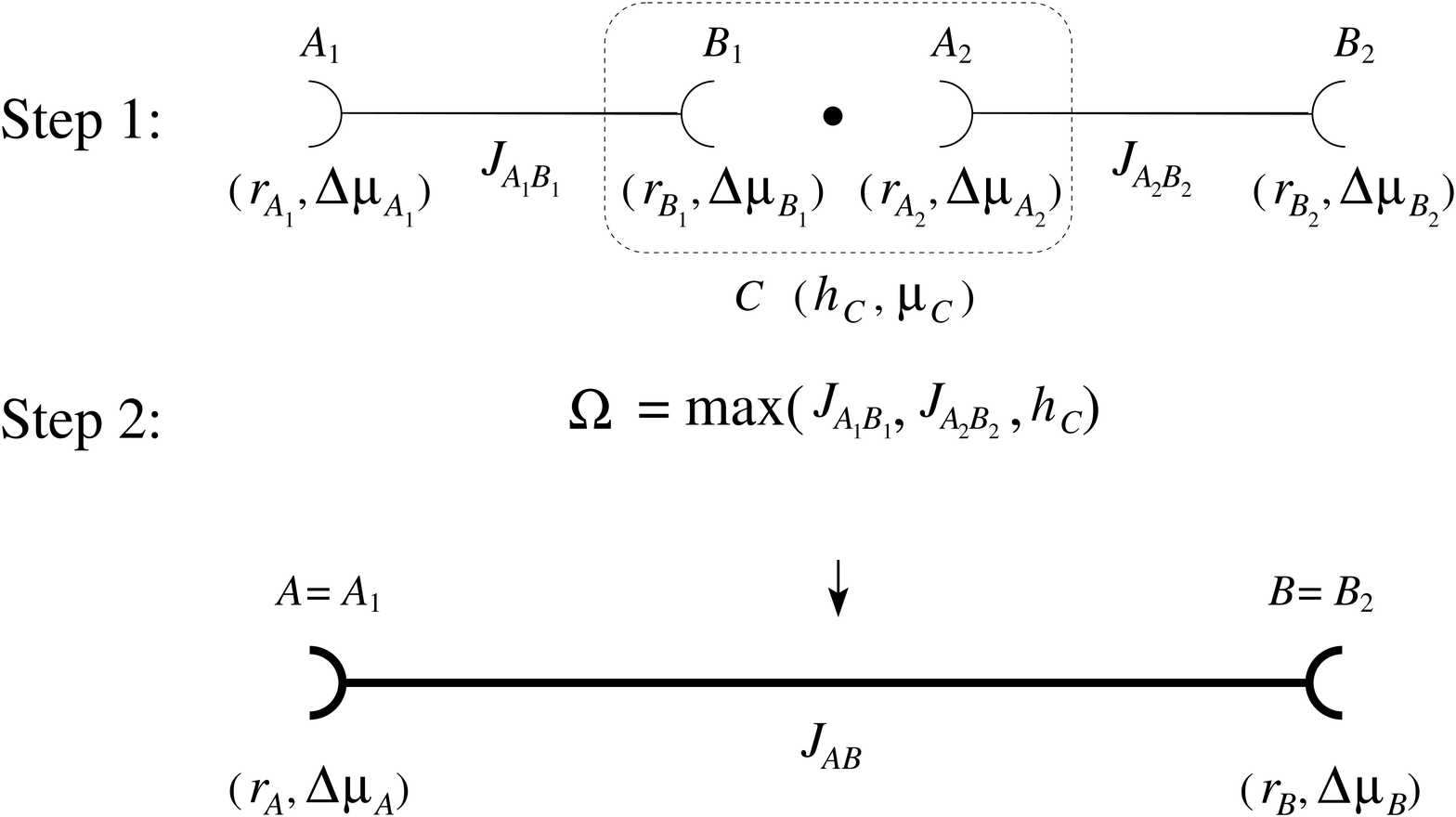}
\caption{ To construct an open bond of generation $(n+1)$ with its renormalized variables 
$ (J_{AB}, r_A,\Delta \mu_A ; r_B,\Delta \mu_B) $ defined in section \ref{sec_notations},
there are two steps : \\
Step 1 : one draws $b$ (here $b=2$) independent open bonds of generation $n$ with their renormalized variables,
 and one connects them via
intermediate points $C_j$ with renormalized variables (see section \ref{sec_step1} for more details). \\
Step 2 : one applies the usual strong disorder RG rules
to the internal structure (see section \ref{sec_step2} for more details). \\
The corresponding RG equation for $b=2$ is written explicitely in Eq. \ref{rgd1}
}
 \label{fig1d}
 \end{figure}

As an example, we now describe more explicitly the case of rescaling factor $b=2$ (see Fig. \ref{fig1d}).
Step 1 consists in drawing $b=2$ independent renormalized open bonds at generation $n$,
with their renormalized variables $(J_{A_1B_1}, r_{A_1},\Delta \mu_{A_1} ; r_{B_1},\Delta \mu_{B_1})$
and $(J_{A_2B_2}, r_{A_2},\Delta \mu_{A_2} ; r_{B_2},\Delta \mu_{B_2})$,
and in the merging of $A_2$ and $B_1$ into a single internal site $C$ 
that has thus for renormalized transverse field
\begin{eqnarray}
h_{C}=h_{C}^{(0)}\ r_{A_{2}} \ r_{B_1}
\label{hCjd1bis}
\end{eqnarray}
and for renormalized magnetic moment
\begin{eqnarray}
\mu_{C}=\mu_{C}^{(0)}+ \Delta \mu_{A_{2}} + \Delta \mu_{B_1}
\label{muCjd1bis}
\end{eqnarray}

Step 2 consists in the Strong Disorder Renormalization
of the internal structure made of the site $C$ and of the two ferromagnetic bonds surrounding $C$,
i.e.  Eq. \ref{omegab} reduces to
\begin{eqnarray}
\Omega= {\rm max } \left[J_{AC}=J_{A_1B_1},h_{C}=h_C^{(0)} r_{A_2} r_{B_1} ,J_{CB}=J_{A_2B_2} \right]
\label{omegab2}
\end{eqnarray}

( i) If $\Omega= h_{C}$, then the site $C$ is decimated according to the usual
Strong Disorder RG rules of Eq A2 : the renormalized coupling reads
\begin{eqnarray}
J_{AB} = \frac{ J_{AC} J_{CB} }{h_{C}}
\label{jnewb2}
\end{eqnarray}
whereas the properties of the two ends are unchanged, i.e. 
the supplementary factors coming from this second step are
\begin{eqnarray}
r_{A}^{(step \ 2)}= 1  \nonumber \\
\Delta \mu_{A}^{(step \ 2)}=0
\label{Anewb2}
\end{eqnarray}

(ii) If $\Omega= J_{AC}$, then the site $C$ is merged with the end $A$
 according to the usual Strong Disorder RG rules of Eqs A3 and A4 : 
the properties of the end $A$ get renormalized according to
\begin{eqnarray}
r_{A}^{(step \ 2)}= \frac{h_{C}}{ J_{AC}  } \nonumber \\
\Delta \mu_{A}^{(step \ 2)}=\mu_{C}
\label{ACb2}
\end{eqnarray}
whereas the renormalized coupling is (Eq A5)
\begin{eqnarray}
J_{AB} =  J_{CB} 
\label{jnewjAb2}
\end{eqnarray}

(iii) If $\Omega= J_{CB}$, then the site $C$ is merged with the end $B$
 according to the usual Strong Disorder RG rules of Eqs A3 and A4: 
the properties of the end $B$ get renormalized according to
\begin{eqnarray}
r_{B}^{(step \ 2)}= \frac{h_{C}}{ J_{C B}  } \nonumber \\
\Delta \mu_{B}^{(step \ 2)} =\mu_{C}
\label{Bnewb2}
\end{eqnarray}
whereas the renormalized coupling is (Eq A5)
\begin{eqnarray}
J_{AB} =  J_{AC} 
\label{jnewjBb2}
\end{eqnarray}

The renormalized open bond of generation $(n+1)$ is thus characterized
by these correlated final variables 
 $(J_{AB}^{final}, r_A^{final},\Delta \mu_A^{final} ; r_B^{final},\Delta \mu_B^{final})$,
where the coupling $J_{AB}^{final}$ has been computed either with (i),(ii),(iii),
and where the variables associated to the end-points $A$ and $B$ contain
the contributions of the two steps described above,
i.e. 
the excess of magnetizations of the two boundaries read
\begin{eqnarray}
\Delta \mu_{A}^{final}=\Delta \mu_{A}^{(step \ 1)} +  \Delta \mu_{A}^{(step \ 2)}  \nonumber \\
\Delta \mu_{B}^{final}=\Delta \mu_{B}^{(step \ 1)} +  \Delta \mu_{B}^{(step \ 2)} 
\label{mustep1et2b2}
\end{eqnarray}
while their multiplicative factors read
\begin{eqnarray}
r_{A}^{final}=r_{A}^{(step \ 1)}  r_{A}^{(step \ 2)} \nonumber \\
r_{B}^{final}=r_{B}^{(step \ 1)}  r_{B}^{(step \ 2)}
\label{rstep1et2b2}
\end{eqnarray}

Equivalently, the joint distribution $P_n(J,r_{A},\Delta \mu_A, r_{B}, \Delta \mu_A)$ evolves
according to the RG equation

\begin{eqnarray}
&& P_{n+1}(J_{AB},r_A, \Delta \mu_A ,r_B,\Delta \mu_B ) =  \int dJ_{A_1B_1} dr_{A_1} d \Delta \mu_{A_1}  d r_{B_1} d \Delta \mu_{B_1}
P_n(J_{A_1B_1},r_{A_1}, \Delta \mu_{A_1}, r_{B_1}, \Delta \mu_{B_1}) \nonumber \\
&& \int dJ_{A_2B_2} dr_{A_2} d \Delta \mu_{A_2} d r_{B_2} d \Delta \mu_{B_2}
P_n(J_{A_2B_2},r_{A_2}, \Delta \mu_{A_2}, r_{B_2}, \Delta \mu_{B_2}) 
 \nonumber \\ && 
\int dh_{C}  d\mu_{C} \delta(h_{C}-h r_{B_1} r_{A_2})  \delta( \mu_{C}- (1+\Delta \mu_{B_1}+ \Delta \mu_{A_2})  \nonumber \\
&& [ \theta(h_{C}>J_{A_1B_1}) \theta(h_{C}>J_{A_2B_2})  \nonumber \\ && 
\delta(J_{AB}-\frac{J_{A_1B_1} J_{A_2B_2})}{h_{C}})
\delta(r_A-r_{A_1}) \delta(\Delta \mu_A-\Delta \mu_{A_1}) \delta(r_B-r_{B_2}) \delta(\Delta \mu_B-\Delta \mu_{B_2})
 \nonumber \\ && 
+ \theta(J_{A_1B_1}>h_{C}) \theta(J_{A_1B_1}>J_{A_2B_2}) \nonumber \\
&& \delta(J_{AB}-J_{A_2B_2})
\delta(r_A-r_{A_1} \frac{h_{C}}{J_{A_1B_1}}) \delta(\Delta \mu_A-(\Delta \mu_{A_1}+\mu_{C}) )
\delta(r_B-r_{B_2}) \delta(\Delta \mu_B-\Delta \mu_{B_2})
 \nonumber \\ && 
 + \theta(J_{A_2B_2}>h_{C}) \theta(J_{A_2B_2}>J_{A_1B_1})  \nonumber \\
&& \delta(J_{AB}-J_{A_1B_1})
\delta(r_A-r_{A_1} ) \delta(\Delta \mu_A-\Delta \mu_{A_1})
\delta(r_B-r_{B_2}\frac{h_{C}}{J_{A_2B_2}})\delta(\Delta \mu_B-(\Delta \mu_{B_2}+\mu_{C}) ] 
\label{rgd1}
\end{eqnarray}
where the two first lines corresponds to the draw of two independent open bonds of generation $n$,
the third line corresponds to the merging of $A_2$ and $B_1$ into the intermediate site $C$,
and where the last lines correspond to the three possible decimations
(respectively of $h_{C}$, of $J_{A_1B_1}=J_{AC} $ and of $J_{A_2B_2} =J_{CB}$).

\subsection{ Discussion  }

We have described in detail the case $d=1$.
It is clear that in the limit of infinite block size $b \to +\infty$, one recovers the full 
usual Strong Disorder RG.
In the next section \ref{sec_nume1d}, we have thus chosen to study numerically
 the 'worst' case of rescaling factor $b=2$
to compare with the exact results of the full Strong Disorder RG corresponding to $b\to +\infty$.
The results with other block sizes $b=3,4,...$ are expected to interpolate between $b=2$ and $b\to +\infty$.

More generally, besides this one-dimensional case,
we can apply the same strategy to other exactly renormalizable lattices.
As discussed in section \ref{sec_order}, the properties of the ordered phase
depend on the existence of an underlying classical ferromagnetic phase.
We have thus chosen to consider below  the two following cases :

(i) in section \ref{sec_triangle}, we consider the Sierpinski gasket
where there is no underlying classical ferromagnetic phase.

(2) in section \ref{sec_b2c8}, we consider the hierarchical diamond Lattice
where there is an underlying classical ferromagnetic phase.

\section{ Numerical results in $d=1$  for the rescaling factor $b=2$ }

\label{sec_nume1d}

We have applied the procedure explained in detail in the previous section
for the rescaling factor $b=2$. In this section, we describe our numerical results
and compare with the exact solution of the full Strong Disorder RG \cite{danieltransverse}.

\subsection{ Numerical details }

We have used the flat initial distribution of couplings on the interval $0 \leq J \leq 1$
\begin{eqnarray}
\pi_{coupling}(J)= \theta(0 \leq J \leq 1)
\label{piJBox}
\end{eqnarray}
and a uniform transverse field $h$.
This choice of box-distribution is very common in numerical studies of random transverse field Ising models
\cite{fisherreview,motrunich,lin,karevski,lin07,yu,kovacsstrip,
kovacs2d,kovacs3d,kovacsentropy,kovacsreview,pich,rieger}.
In addition in dimension $d=1$, it is exactly known that any infinitesimal disorder 
flows towards the same Infinite-Disorder fixed point \cite{danieltransverse}.

The corresponding exact transition point given
 by the criterion $\overline{ \ln h} = \overline{ \ln J}$ \cite{danieltransverse}
is
\begin{eqnarray}
h_c^{exact} = e^{ \overline{\ln J}} = e^{-1}=0.36787944...
\label{hcexactd1}
\end{eqnarray}

We have performed numerical simulations with the so-called 'pool-method'
which is very much used for disordered systems on hierarchical lattices,
in particular for spin-glasses \cite{hierarchicalspinglass},
but also for disordered polymers \cite{Coo_Der,diamondcriti}.
The idea is to represent the joint probability distribution
$P_n(J,r_{A},\Delta \mu_A, r_{B}, \Delta \mu_A)$  at generation
$n$, by a pool of $M$ realizations $ (J^{(i)},r_{A}^{(i)},\Delta \mu_A^{(i)}, r_{B}^{(i)}, \Delta \mu_A^{(i)})$
where $i=1,2,...,M$.
The pool at generation $(n+1)$ is then obtained as follows :
each new set  $ (J^{(i)},r_{A}^{(i)},\Delta \mu_A^{(i)}, r_{B}^{(i)}, \Delta \mu_A^{(i)})$
of generation $(n+1)$  is obtained by choosing 
$b=2$ values $i$ at random with their corresponding variables in the pool of generation $n$.

The results presented in this Section have been obtained
with a pool of size $M=10^7$. We find that the corresponding transition point satisfies
\begin{eqnarray}
0.367910 < h_c^{pool}(M=10^7) < 0.367912
\label{hcpoold1}
\end{eqnarray}
i.e. it is slightly larger than the exact value of Eq. \ref{hcexactd1} as a consequence
of the finite size $M$ of the pool.
We now describe our numerical results concerning
the RG flow of the renormalized variables in various phases using $0 \leq n \leq 50$ generations
corresponding to lengths 
\begin{eqnarray}
1 \leq L=2^n \leq 2^{50}
\label{sizesd1}
\end{eqnarray}

\subsection{ RG flow of the typical renormalized coupling $J_L^{typ}$ }

\begin{figure}[htbp]
%\begin{figure}
 \includegraphics[height=6cm]{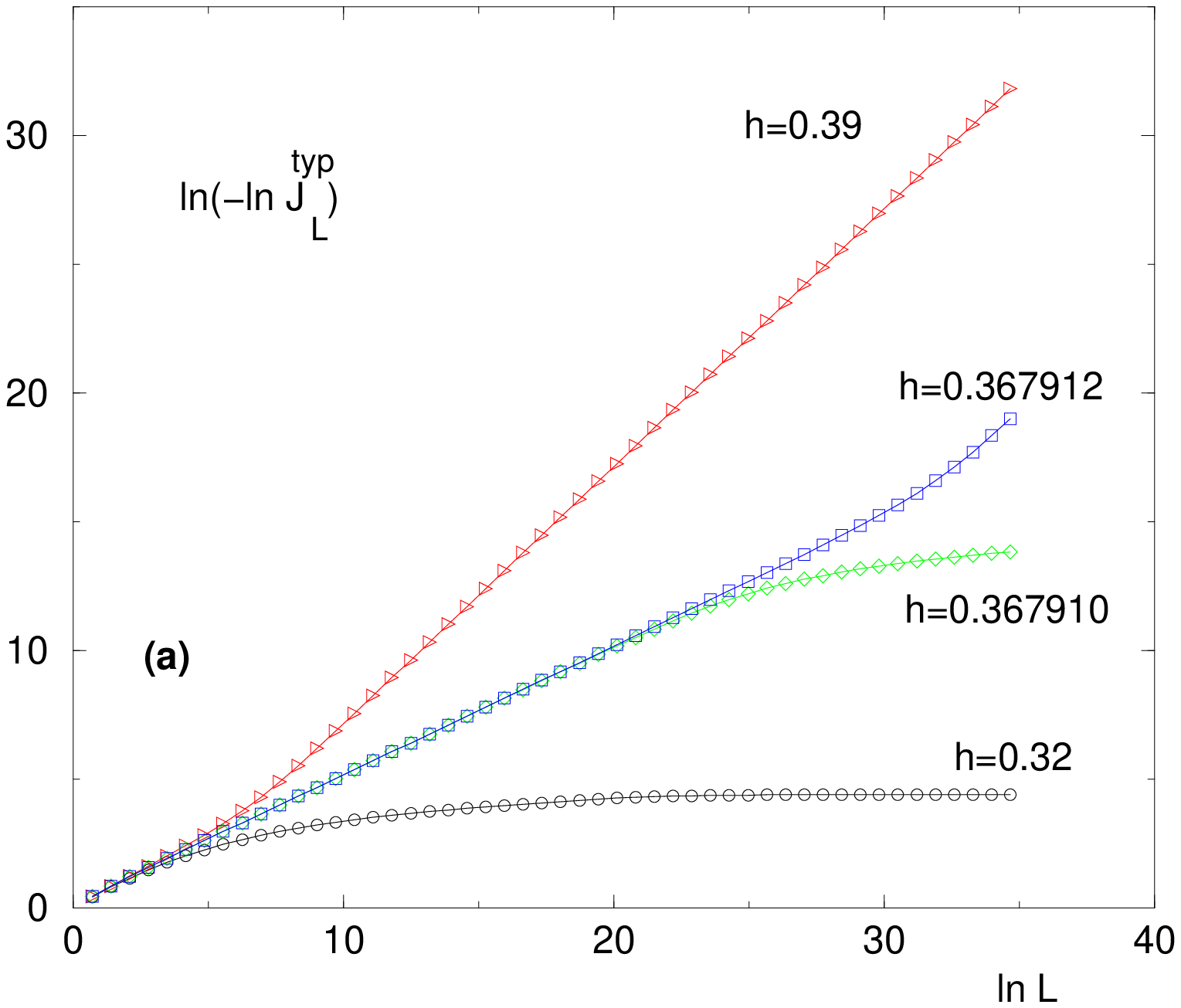}
\hspace{1cm}
 \includegraphics[height=6cm]{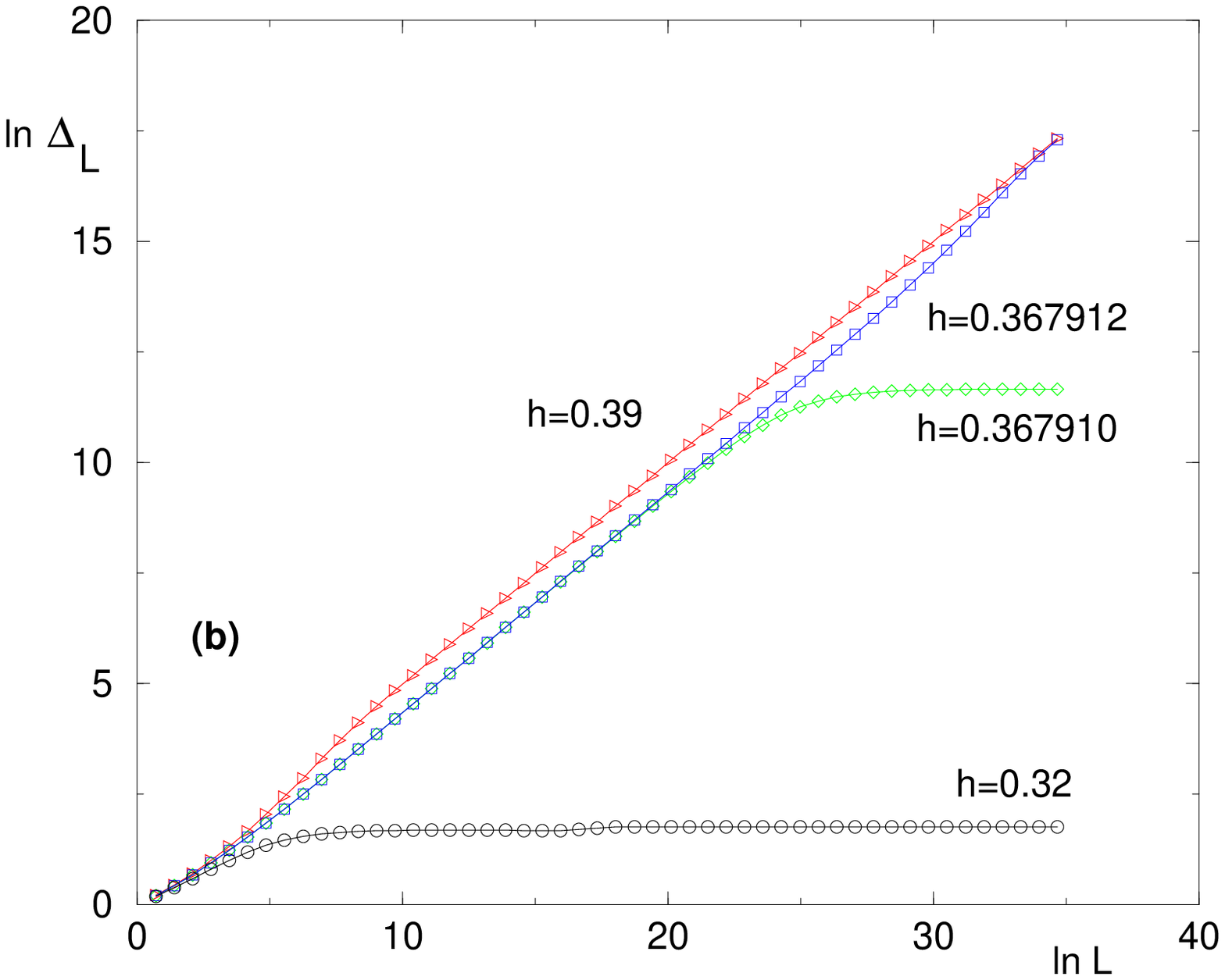}
\caption{ ($d=1$ ) 
(a)  RG flow of the logarithm of the typical renormalized coupling $(\ln J_L^{typ})$ in a log-log plot :
  the slope is $0$ in the ordered phase (e.g. $h=0.32$),
  the slope is $1$ in the disordered phase (e.g. $h=0.39$),
and the slope is $\psi \simeq 0.5 $ at criticality (before the bifurcation 
of the curves corresponding to $h=0.367910$ and $h=0.367912$ ).
(b)  RG flow of the width $\Delta_L$ of the distribution of the logarithm of the renormalized couplings in a log-log plot : the slope is $0$ in the ordered phase (e.g. $h=0.32$),
the slope is $\omega \simeq 0.5$ in the disordered phase (e.g. $h=0.39$),
and the slope is $\psi \simeq 0.5$ at criticality (before the bifurcation 
of the curves corresponding to $h=0.367910$ and $h=0.367912$ ).
  }
\label{fig1drgflowj}
\end{figure}

On Fig. \ref{fig1drgflowj} (a), we show our data concerning the RG flow of the typical coupling 
\begin{eqnarray}
J_L^{typ}=e^{\overline{\ln J_L }}
\label{defjtyp}
\end{eqnarray}
Our results for the $L$-dependence in the two phases and at criticality
\begin{eqnarray}
\ln J_L^{typ} \vert_{h<h_c} && \oppropto_{L \to +\infty} Cst  \nonumber \\
\ln J_L^{typ} \vert_{h=h_c}  && \oppropto_{L \to +\infty} - L^{\psi} \ \ {\rm with } \ \ \psi \simeq 0.5  \nonumber \\
\ln J_L^{typ} \vert_{h>h_c} && \oppropto_{L \to +\infty} -L 
\label{jtyp1d}
\end{eqnarray}
are in agreement with the exact solution \cite{danieltransverse}.

\subsection{ RG flow of the width $\Delta_L$ of the logarithms of the renormalized couplings }

On Fig. \ref{fig1drgflowj} (b), we show our data concerning the RG flow of the width $\Delta_L$
 of the distribution of the logarithms of the couplings.
Again, our results for the $L$-dependence in the two phases and at criticality
\begin{eqnarray}
\Delta_L \vert_{h<h_c}  && \oppropto_{L \to +\infty} Cst  \nonumber \\
\Delta_L \vert_{h=h_c}  && \oppropto_{L \to +\infty}  L^{\psi} \ \ {\rm with } \ \ \psi \simeq 0.5  \nonumber \\
\Delta_L \vert_{h>h_c} && \oppropto_{L \to +\infty} L^{\omega} \ \ {\rm with } \ \ \omega \simeq 0.5
\label{deltaL1d}
\end{eqnarray}
are in agreement with the exact solution \cite{danieltransverse}.

\begin{figure}[htbp]
%\begin{figure}
 \includegraphics[height=6cm]{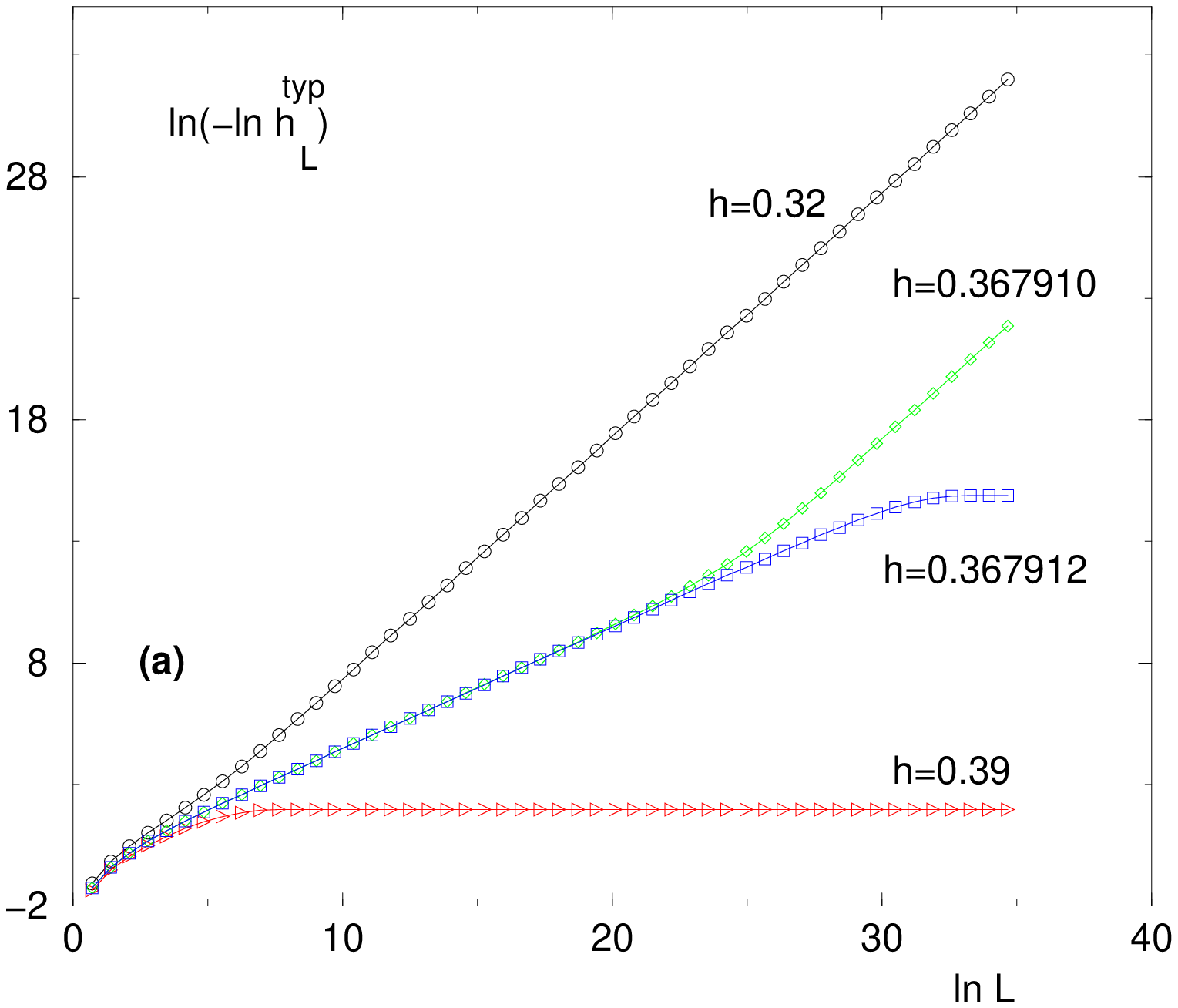}
\hspace{1cm}
 \includegraphics[height=6cm]{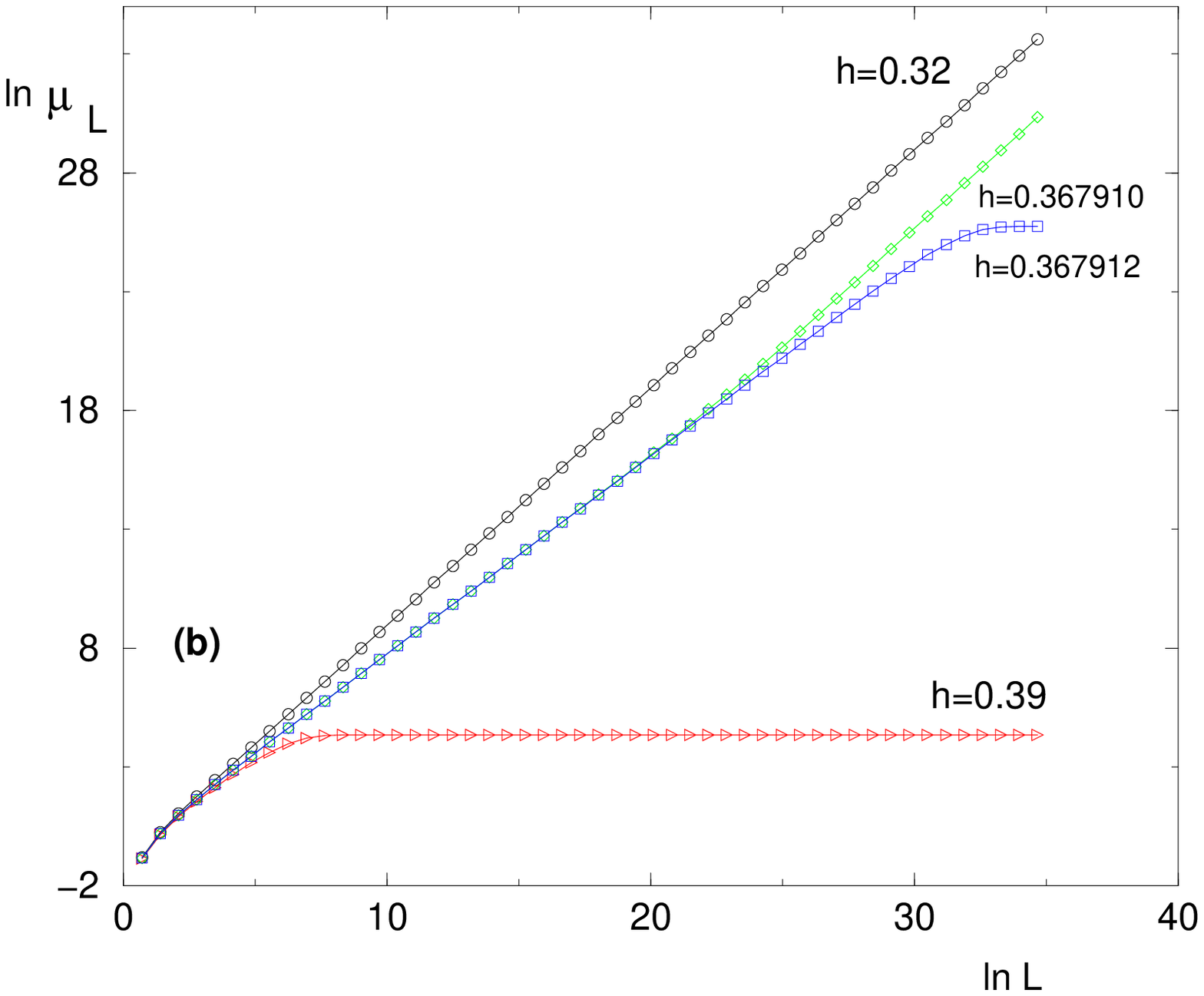}
\caption{ ( $d=1$ ) 
(a)  RG flow of the logarithm of the typical renormalized transverse field $(\ln h_L^{typ})$   in a log-log plot :
  the slope is $d=1$ in the ordered phase (e.g. $h=0.32$),
  the slope is $0$ in the disordered phase (e.g. $h=0.39$),
and the slope is $\psi \simeq 0.5 $ at criticality (before the bifurcation 
of the curves corresponding to $h=0.367910$ and $h=0.367912$ ).
(b)  RG flow of the renormalized magnetization $\mu_L$ in a log-log plot : 
the slope is $d=1$ in the ordered phase (e.g. $h=0.32$),
the slope is $0$ in the disordered phase (e.g. $h=0.39$),
and the slope is $d_f \simeq 0.85$ at criticality (before the bifurcation 
of the curves corresponding to $h=0.367910$ and $h=0.367912$).
  }
\label{fig1drgflowmuh}
\end{figure}

\subsection{ RG flow of the typical renormalized transverse field $h_L^{typ}$}

On Fig. \ref{fig1drgflowmuh} (a), we show the RG flow of the typical renormalized transverse field
\begin{eqnarray}
h_L^{typ}=e^{\overline{\ln h_L }}
\label{defhtyp}
\end{eqnarray}
Again, our results for the $L$-dependence in the two phases and at criticality
\begin{eqnarray}
\ln h_L^{typ} \vert_{h<h_c} && \oppropto_{L \to +\infty} -L  \nonumber \\
\ln h_L^{typ} \vert_{h=h_c}  && \oppropto_{L \to +\infty} - L^{\psi} \ \ {\rm with } \ \ \psi \simeq 0.5  \nonumber \\
\ln h_L^{typ} \vert_{h>h_c} && \oppropto_{L \to +\infty} Cst 
\label{htyp1d}
\end{eqnarray}
are in agreement with the exact solution \cite{danieltransverse}.

\subsection{ RG flow of the renormalized magnetization $\mu_L$ }

On Fig. \ref{fig1drgflowmuh} (b), we show the RG flow of the magnetization $ \mu_L$
of renormalized clusters as a function of $L$.
 In the ordered phase, the magnetization is extensive $\mu_L \propto L$,
whereas in the disordered phase, the magnetization of clusters remain finite  $\mu_L \propto Cst$ as it should.
 At criticality, we measure $\mu_L \propto L^{d_f}$ where
the fractal dimension of critical ordered clusters is of order $d_f \simeq 0.85$
whereas the exact result is 
\begin{eqnarray}
d_f^{exact}= \frac{1+\sqrt{5}}{4}= 0.809...
\label{dfexact1d}
\end{eqnarray}
This discrepancy will be discussed after Eq. \ref{betaexact1d}.

\subsection{ Critical exponents in the disordered phase $h>h_c$ }

\begin{figure}[htbp]
%\begin{figure}
 \includegraphics[height=6cm]{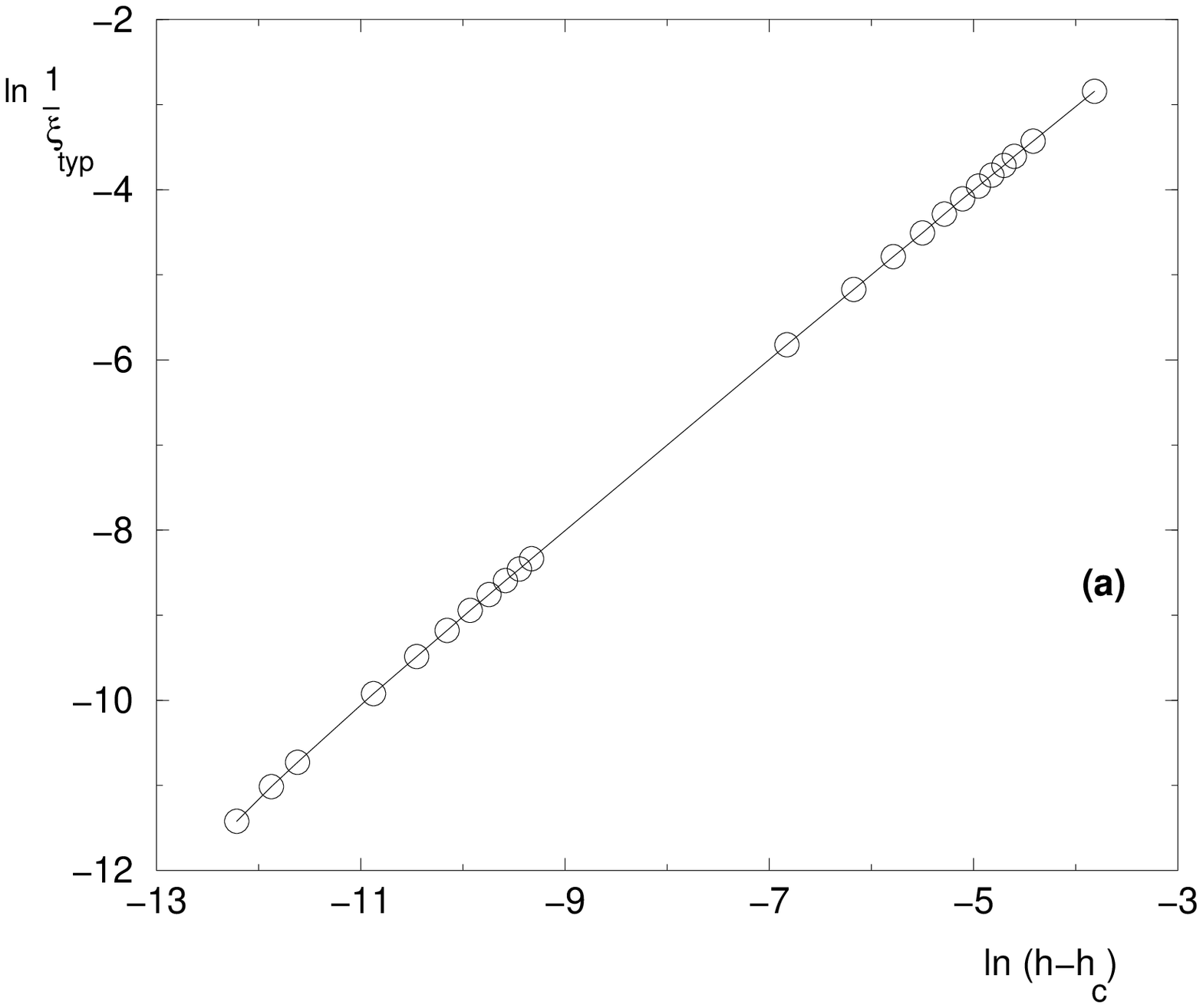}
\hspace{1cm}
\includegraphics[height=6cm]{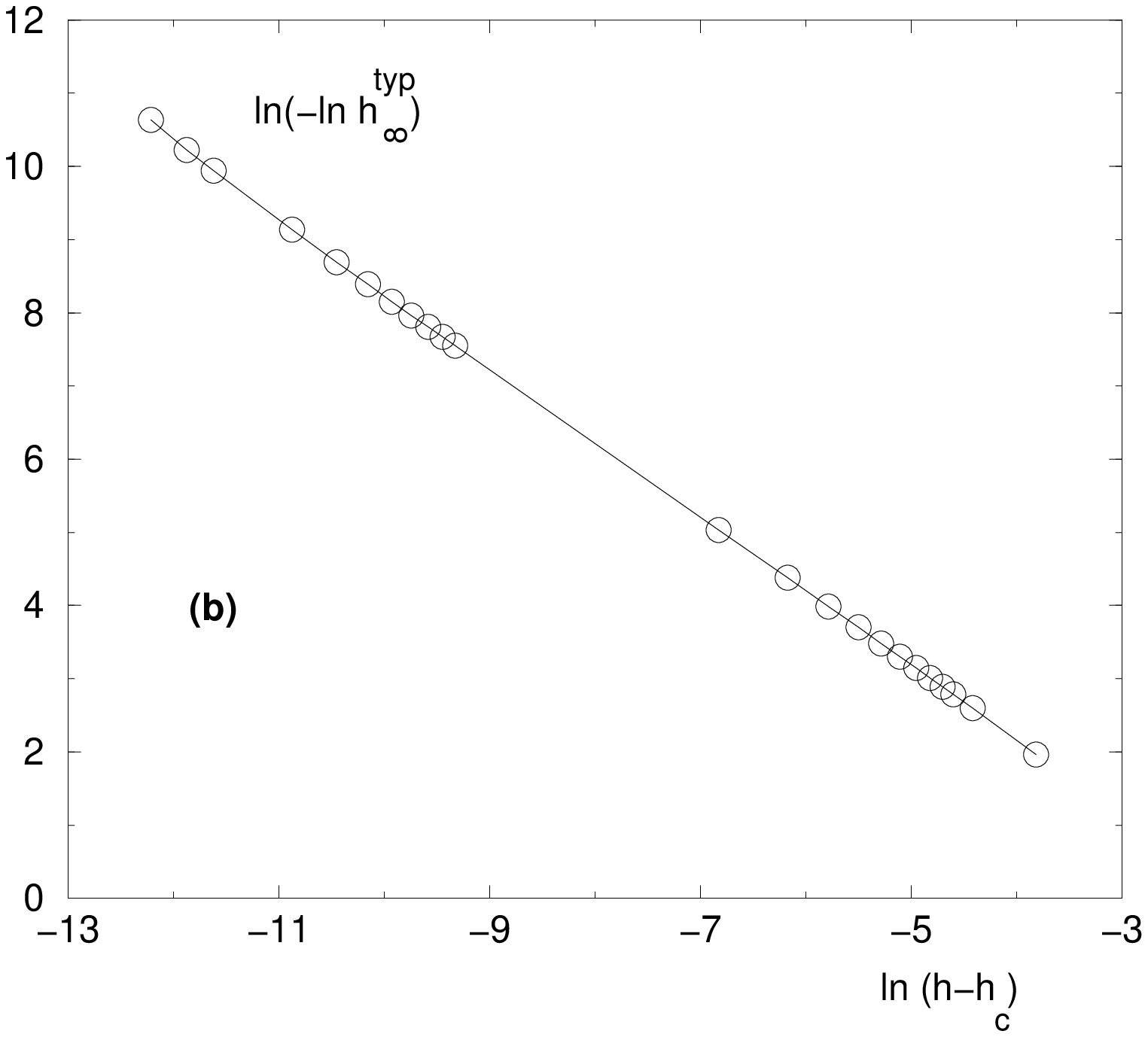}
\caption{ ($d=1$ ) Critical exponents in the disordered phase $h>h_c$ : 
(a) Divergence of the typical correlation length $\xi_{typ}$ of Eq. \ref{xitypdes} :
 we measure $\nu_{typ} \simeq 1$
 (b) Essential singularity of the typical finite renormalized transverse field $h_{\infty}^{typ}$ of Eq. \ref{defkappa} :
 we measure $\kappa \simeq 1$. }
\label{fig1dxitypdes}
\end{figure}

In the disordered phase, the exponential decay of the typical renormalized coupling
$J_L^{typ}=e^{\overline{\ln J_L }}$ defines the typical correlation length $\xi_{typ}$ (Eq \ref{JLdisordertyp})
\begin{eqnarray}
\ln J_L^{typ} \equiv \overline{\ln J_L } \opsimeq_{L \to +\infty} - \frac{L}{\xi_{typ}} 
\label{xitypdes}
\end{eqnarray}
As shown on Fig. \ref{fig1dxitypdes} (a), the divergence at criticality
\begin{eqnarray}
\xi_{typ} \propto (h-h_c)^{-\nu_{typ}}
\label{nutypdes}
\end{eqnarray}
is governed by the exponent
\begin{eqnarray}
\nu_{typ} \simeq 1
\label{nutypdesexact}
\end{eqnarray}
in agreement with the exact solution \cite{danieltransverse}.

As shown on Fig. \ref{fig1dxitypdes} (b), the finite typical renormalized transverse field 
$h_{\infty}^{typ}$ displays the essential singularity (Eq. \ref{defkappa}) with
\begin{eqnarray}
\kappa \simeq 1
\label{kappa1d}
\end{eqnarray}
 in agreement with the exact solution \cite{danieltransverse}.
Both values of Eq. \ref{nutypdesexact} and Eq. \ref{kappa1d} correspond to
the finite-size correlation exponent (see Eqs \ref{kappafss} and \ref{nutyp})
\begin{eqnarray}
\nu_{FS} \simeq 2
\label{nufs1d}
\end{eqnarray}
 in agreement with the exact solution \cite{danieltransverse}.

\subsection{ Critical exponents in the ordered phase $h<h_c$ }

\begin{figure}[htbp]
% \includegraphics[height=6cm]{xm1dmagnetisation84.eps}
% \hspace{1cm}
 \includegraphics[height=6cm]{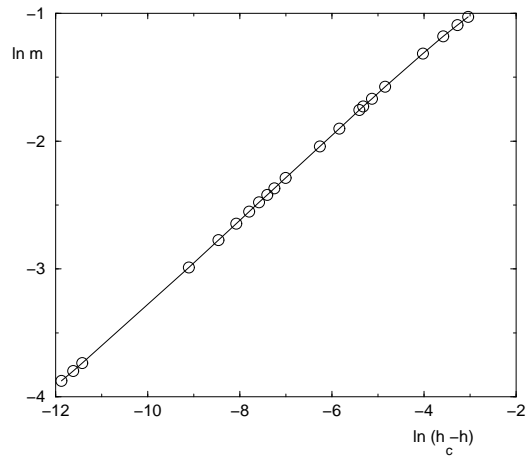}
\caption{ ($d=1$ ) Singularity of the intensive magnetization $m$ in the ordered phase 
(Eq. \ref{defbeta}) :
we measure an exponent of order $\beta \simeq 0.33$ instead 
of the exact value of Eq. \ref{betaexact1d} (see explanations in the text).
  }
\label{fig1dmagne}
\end{figure}

In the ordered phase,  
the exponential decay of the typical renormalized transverse field
 defines the length $\xi_h$ (Eq. \ref{hLorder})
\begin{eqnarray}
\ln  h_L^{typ}\equiv \overline{\ln h_L} \opsimeq_{L \to +\infty} - \frac{L}{\xi_h} 
\label{xityporder1d}
\end{eqnarray}
where $\xi_h$ diverges with the exponent $\nu_{h} \simeq 1$
in agreement with the exact solution \cite{danieltransverse}.
Similarly, the finite typical renormalized coupling
$J_{\infty}^{typ}$ displays an essential singularity 
with the same exponent $\kappa \simeq 1$ as in Eq. \ref{kappa1d},
in agreement with the exact solution \cite{danieltransverse}.
On Fig. \ref{fig1dmagne},
 we show our data concerning the intensive magnetization of Eq. \ref{defbeta}
\begin{eqnarray}
\overline{m} = \frac{ \overline{ \mu_L} } { L } \propto (h_c-h)^{\beta}
\label{magne1d}
\end{eqnarray}
We measure an exponent of order $\beta \simeq 0.33$
instead of the exact result
\begin{eqnarray}
\beta^{exact}=\frac{3-\sqrt{5}}{2}= 0.38195...
\label{betaexact1d}
\end{eqnarray}
We believe that this discrepancy, related to the previous discrepancy found at criticality 
(see Eq \ref{dfexact1d}) comes from the rare-event nature of the magnetization
in the critical region : indeed, the critical magnetization 
 represents some 'persistence exponent' 
for the strong disorder renormalization flow \cite{hastings}, 
because it is related to the probability for a given spin
to remain undecimated during the RG flow \cite{danieltransverse}.
Within our framework where some spins are declared 'undecimable' up to some given stage
of the real space RG, this persistence probability is changed 
so that the corresponding exponents for the magnetization are not captured exactly
by our fixed-cell-size procedure.
Nevertheless, the approximated exponents obtained are expected to become better and better
as the rescaling factor $b$ grows, and to converge to the exact values as $b \to +\infty$

\subsection{ Discussion} 

In summary, except for the magnetic exponent $\beta$ and related exponents like $d_f$
that are not reproduced exactly for finite $b$ (for reasons explained just above),
the fixed cell-size procedure with the rescaling factor $b=2$ is able to capture
correctly the other critical exponents, in particular the activated exponent $\psi$,
    the typical correlation exponent $\nu_{typ}$, and the essential singularity exponent $\kappa=1$.
For other rescaling factors $b=3,4,..$, we expect that these critical exponents
 $(\psi,\nu_{typ},\kappa)$ will be again exactly reproduced (since $b=2$ is the 'worst'
approximation with respect to the full rules corresponding to $b\to +\infty$),
whereas the magnetic exponents $\beta(b)$ will be better approximations of the true exponent
$\beta_{exact}=\beta(b \to +\infty)$ than $\beta(b=2)$, but will never be exact for 
any finite $b$.
We also believe that these conclusions can be extended to more complicated observables
like spin-spin correlations as follows : typical observables (like typical correlations
that involve the exponent $\psi$ at criticality) should be reproduced with the correct critical exponents,
whereas averaged observables governed by rare persistent events (like averaged correlations
that involve the magnetic exponent $\beta$) cannot be correctly reproduced for any finite $b$.

After having checked that typical exponents can be correctly reproduced in $d=1$ with $b=2$,
 we may now apply this fixed cell-size procedure to fractal lattices of dimension $d>1$.

\section{ Numerical results for the Sierpinski gasket }

\label{sec_triangle}

The Sierpinski gasket is one of the simplest hierarchical lattice made of nested triangles.
It has for fractal dimension
\begin{eqnarray}
D= \frac{\ln 3 }{\ln 2 } = 1.5849525...
\label{dftriangle}
\end{eqnarray}
There is no underlying ferromagnetic phase (i.e. $T_c=0$), because
the so-called 'ramification' remains finite \cite{gefen}. 

\subsection{ Principle of the fixed cell-size RG procedure }

\begin{figure}[htbp]
\includegraphics[height=6cm]{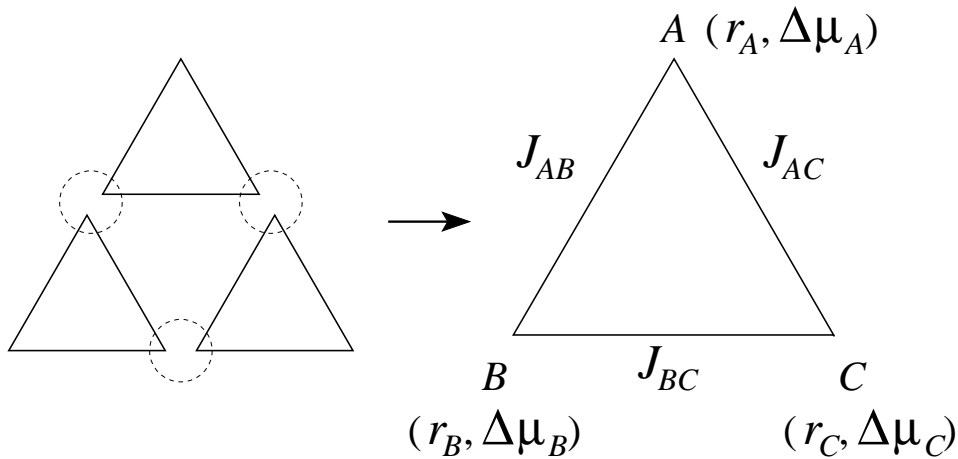}
\caption{ To construct an open triangle of generation $(n+1)$ with its renormalized variables 
$(J_{AB},J_{AC},J_{BC}, r_A, \Delta \mu_A , r_B, \Delta \mu_B,
r_C, \Delta \mu_C)$,
the are two steps : \\
Step 1 : one draw three independent open triangles of generation $n$, and one connects them via
intermediate points  \\
Step 2 : one applies the usual strong disorder RG rules
to the internal structure.
}
 \label{figtriangle}
 \end{figure}

Since the Sierpinski gasket can be constructed recursively from triangles,
it is clear that the fixed cell-size RG procedure based on Strong Disorder RG rules
will be based on the joint probability $P_n(J_{AB},J_{AC},J_{BC}, r_A, \Delta \mu_A , r_B, \Delta \mu_B,
r_C, \Delta \mu_C)$ of 'open triangles',
that will replace the joint probability $P_n( J_{AB}, r_A,\Delta \mu_A ; r_B,\Delta \mu_B)$
of 'open bonds' described in section \ref{sec_notations} :
the variables for one open triangle $ABC$ are 
 the three ferromagnetic couplings $(J_{AB},J_{AC},J_{BC})$ associated to the three bonds of the triangle,
 the three magnetization-excesses $(\Delta \mu_A ,\Delta \mu_B, \Delta \mu_C  )$ associated to the three vertices,
and the three multiplicative factors $(r_A ,r_B, r_C  )$ for the transverse fields of the three vertices.

It is clear that with three open triangles of generation $n$, we can build the structure
of generation $(n+1)$ by following the same principles as in section \ref{sec_step1}.
We may then apply Strong Disorder RG rules to this structure to obtain
an open triangle of generation $(n+1)$ with its renormalized variables,
by following the same principles as in section \ref{sec_step2}
(the only novelty is that in the ordered phase, one of the three renormalized ferromagnetic
coupling of the triangle may vanish).

As in the previous section, we have used the flat initial distribution of Eq. \ref{piJBox}
for the random couplings and a uniform transverse field $h$.
The results presented in this Section have been obtained
with a pool of size $M=7.10^6$, leading to a transition point of order
\begin{eqnarray}
 1.3527037< h_c^{pool}(M=7.10^6) < 1.3527038
\label{hcpooltriangle}
\end{eqnarray}
using $0 \leq n \leq 100$ generations
corresponding to lengths 
\begin{eqnarray}
1 \leq L=2^n \leq 2^{100}
\label{sizestriangle}
\end{eqnarray}

\subsection{ RG flow of the typical renormalized coupling $J_L^{typ}$ }

\begin{figure}[htbp]
%\begin{figure}
 \includegraphics[height=6cm]{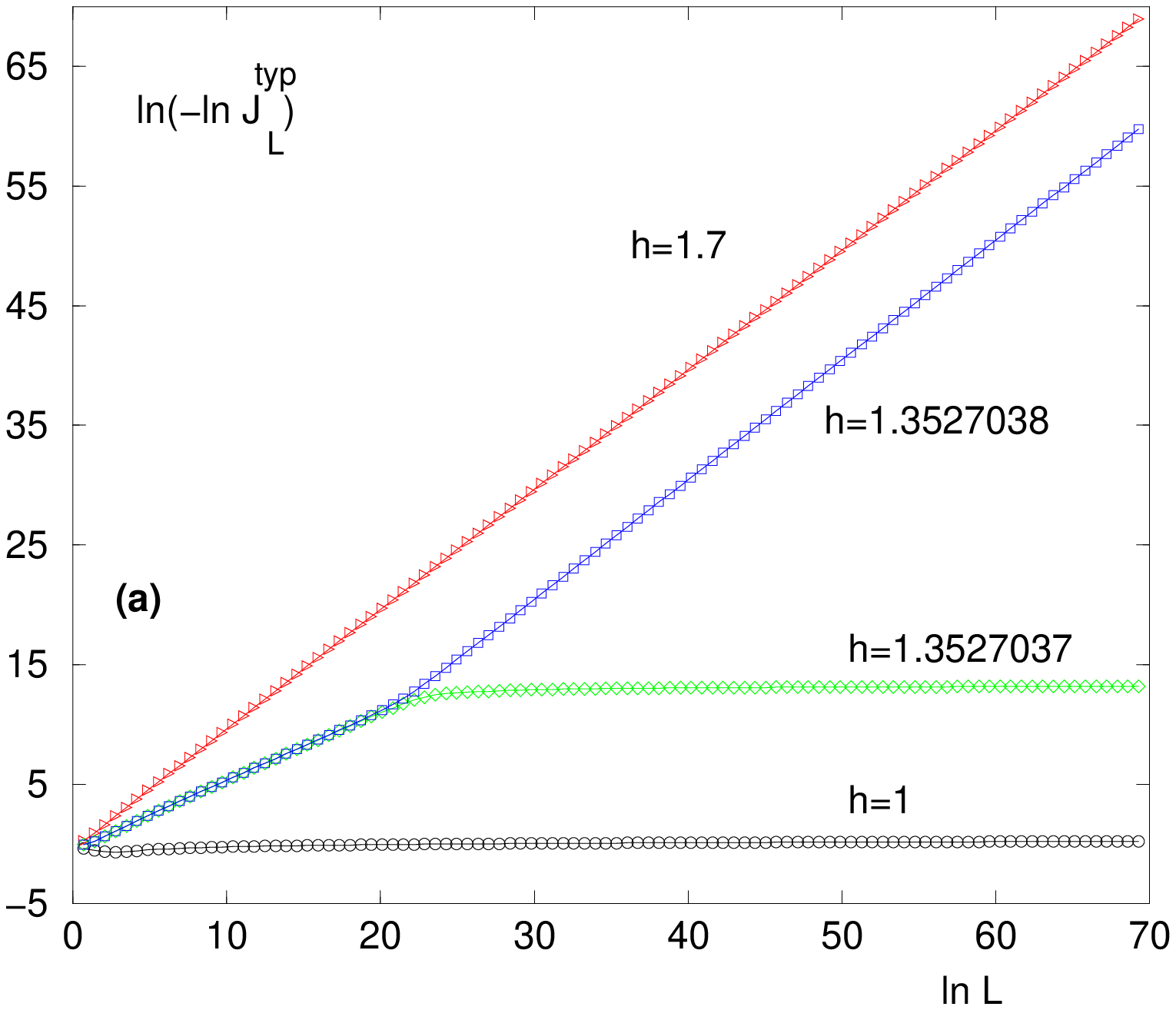}
\hspace{1cm}
 \includegraphics[height=6cm]{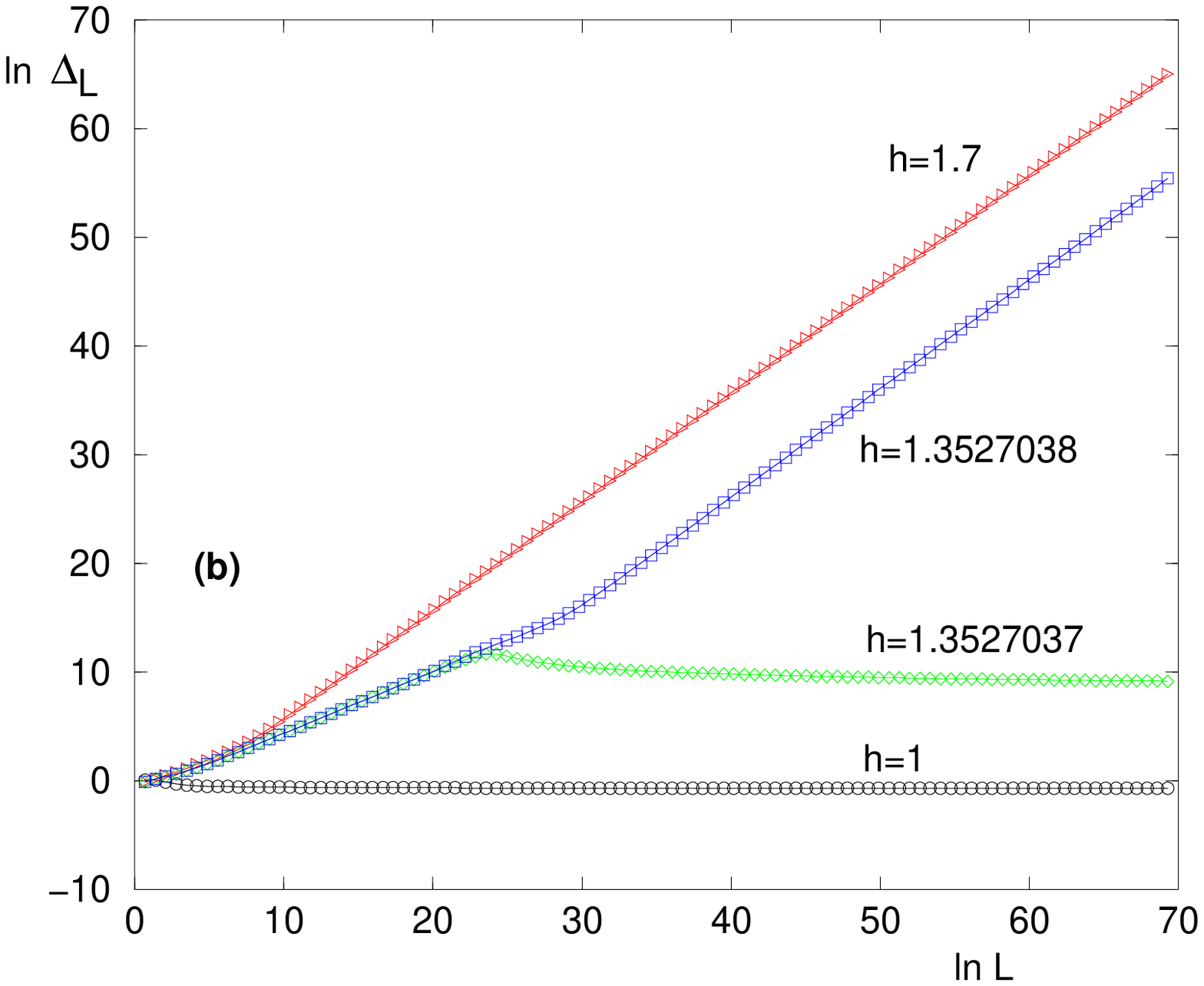}
\caption{ (Sierpinski ) 
(a)  RG flow of the logarithm of the typical renormalized coupling $(\ln J_L^{typ})$ in a log-log plot :
  the slope is $0$ in the ordered phase (e.g. $h=1$),
  the slope is $1$ in the disordered phase (e.g. $h=1.7$),
and the slope is $\psi \simeq 0.58 $ at criticality (before the bifurcation 
of the curves corresponding to $h=1.3527037$ and $h=1.3527038$).
(b)  RG flow of the width $\Delta_L$ of the distribution of the logarithm of the
renormalized couplings in a log-log plot : the slope is $0$ in the ordered phase (e.g. $h=1$),
the slope is $\omega \simeq 1$ in the disordered phase (e.g. $h=1.7$),
and the slope is $\psi \simeq 0.58$ at criticality (before the bifurcation 
of the curves corresponding to $h=1.3527037$ and $h=1.3527038$ ).
  }
\label{figtriangleflowj}
\end{figure}

On Fig. \ref{figtriangleflowj} (a), we show the RG flow of the typical coupling 
$J_L^{typ}=e^{\overline{\ln J_L }}$ as a function of $L$
\begin{eqnarray}
\ln J_L^{typ} \vert_{h<h_c} && \oppropto_{L \to +\infty} Cst  \nonumber \\
\ln J_L^{typ} \vert_{h=h_c}  && \oppropto_{L \to +\infty} - L^{\psi} \ \ {\rm with } \ \ \psi \simeq 0.58  \nonumber \\
\ln J_L^{typ} \vert_{h>h_c} && \oppropto_{L \to +\infty} -L 
\label{jtyptriangle}
\end{eqnarray}

\subsection{ RG flow of the width $\Delta_L$ of the logarithms of the renormalized couplings }

On Fig. \ref{figtriangleflowj} (b), we show the RG flow of  the width $\Delta_L$
 of the distribution of the logarithms of the couplings :

\begin{eqnarray}
\Delta_L \vert_{h<h_c} && \oppropto_{L \to +\infty} Cst  \nonumber \\
\Delta_L  \vert_{h=h_c} && \oppropto_{L \to +\infty}  L^{\psi} \ \ {\rm with } \ \ \psi \simeq 0.58  \nonumber \\
\Delta_L  \vert_{h>h_c}  && \oppropto_{L \to +\infty} L^{\omega} \ \ {\rm with } \ \ \omega \simeq 1
\label{deltaLtriangle}
\end{eqnarray}
Note that the value $\omega=1$ is an 'anomalously' large exponent for
the droplet exponent of the Directed Polymer if one compares with
 hypercubic lattices where $\omega<1$ (Eq. \ref{JLdisorder}).
We believe that this anomaly comes from the hierarchical structure of the Sierpinski gasket
where the polymer is forced to visit certain points at each scale.

\begin{figure}[htbp]
%\begin{figure}
 \includegraphics[height=6cm]{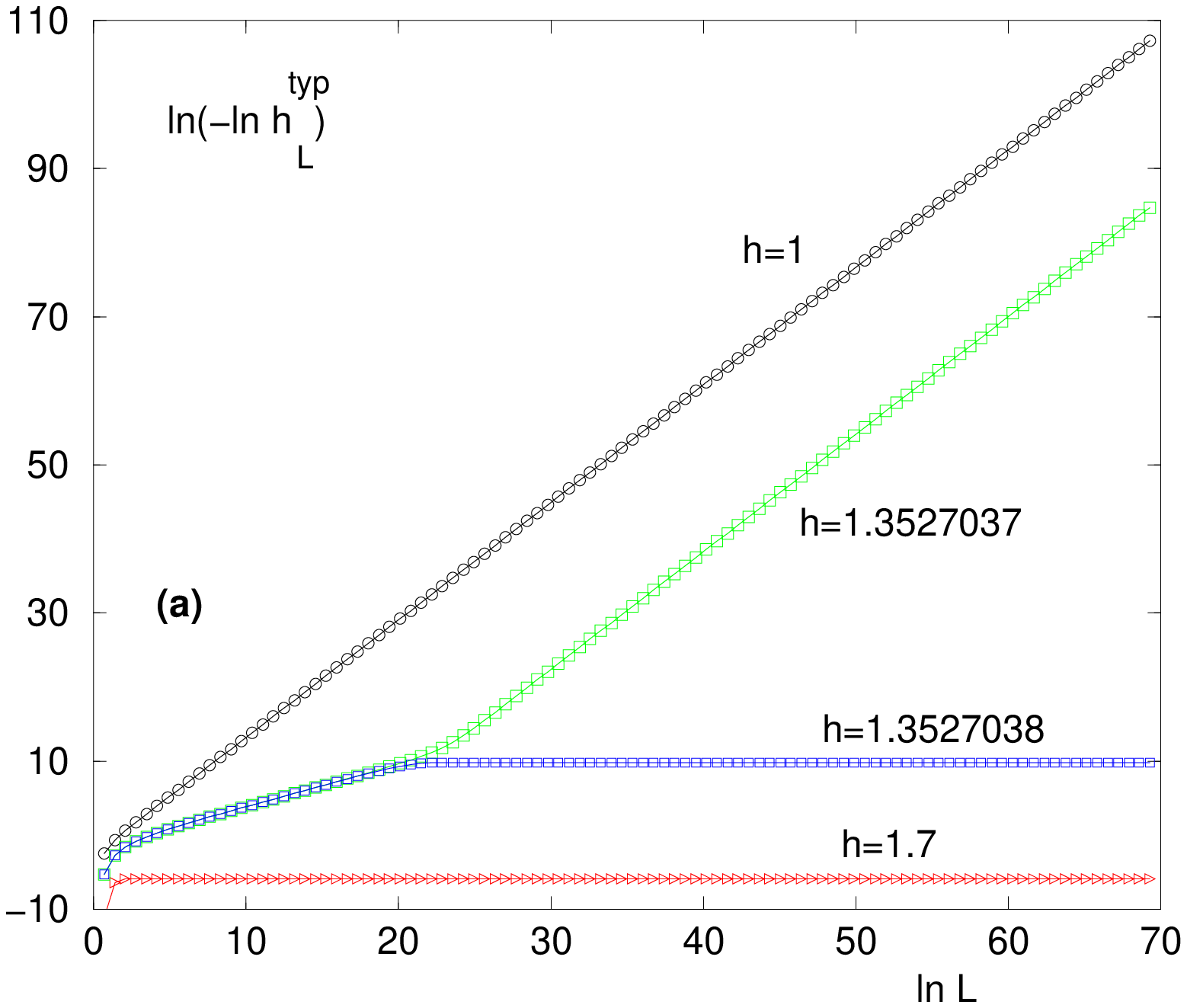}
\hspace{1cm}
 \includegraphics[height=6cm]{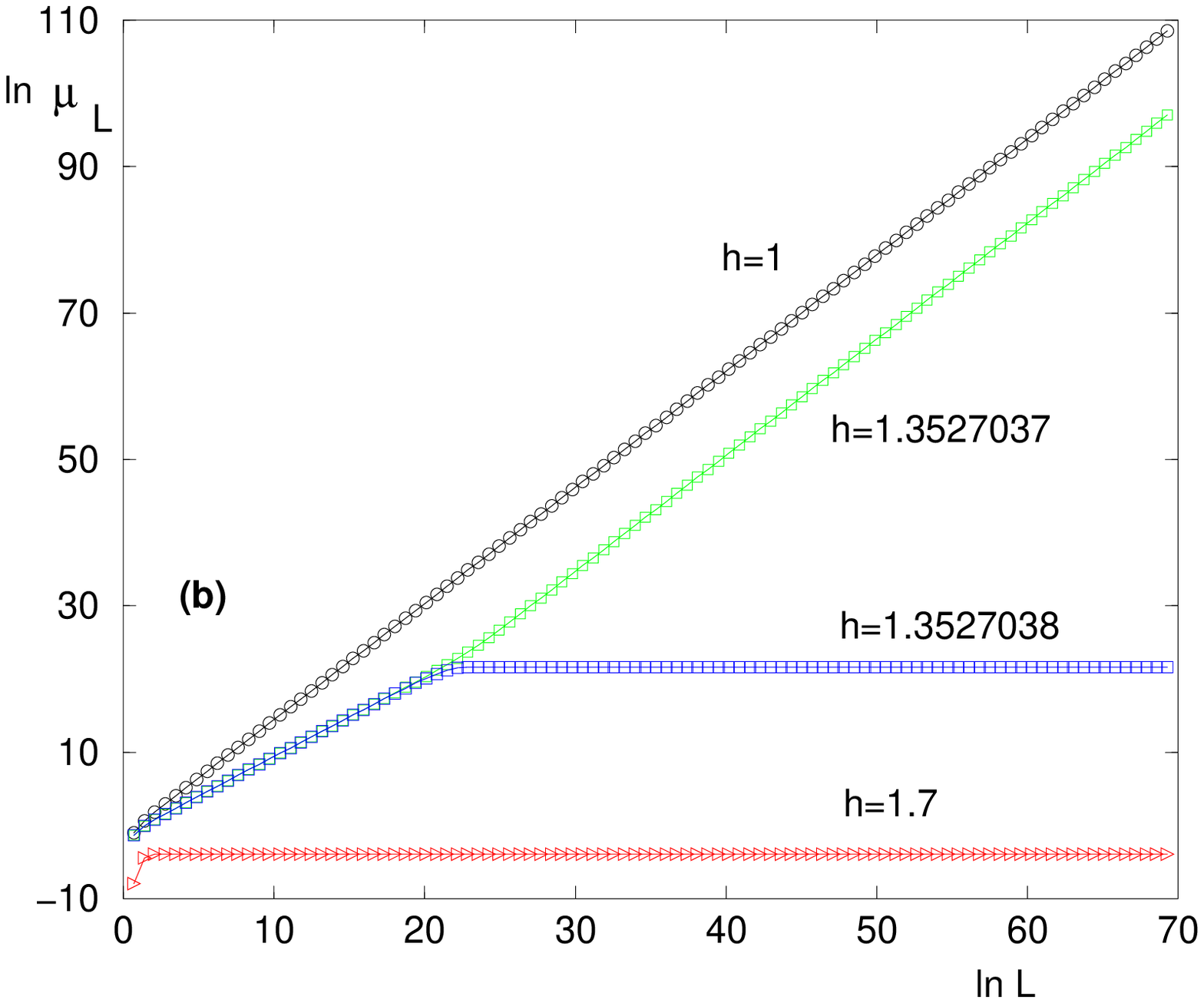}
\caption{ (Sierpinski ) 
(a)  RG flow of the logarithm of the typical renormalized transverse field $(\ln h_L^{typ})$   in a log-log plot :
  the slope is $D=\frac{\ln 3 }{\ln 2 } = 1.58..$ in the ordered phase (e.g. $h=1$),
  the slope is $0$ in the disordered phase (e.g. $h=1.7$),
and the slope is $\psi \simeq 0.58 $ at criticality (before the bifurcation 
of the curves corresponding to $h=1.3527037$ and $h=1.3527038$ ).
(b)  RG flow of the renormalized magnetization $\mu_L$ in a log-log plot : 
the slope is $D=\frac{\ln 3 }{\ln 2 } = 1.58..$   in the ordered phase (e.g. $h=1$),
the slope is $0$ in the disordered phase (e.g. $h=1.7$),
and the slope is $d_f \simeq 1.08$ at criticality (before the bifurcation 
of the curves corresponding to $h=1.3527037$ and $h=1.3527038$).
  }
\label{figtriangleflowmuh}
\end{figure}

\subsection{ RG flow of the typical renormalized transverse field $h_L^{typ}$}

On Fig. \ref{figtriangleflowmuh} (a), we show the RG flow of 
the typical renormalized transverse field $h_L^{typ}$ as a function of $L$
\begin{eqnarray}
\ln h_L^{typ} \vert_{h<h_c}  && \oppropto_{L \to +\infty} -L^D  \nonumber \\
\ln h_L^{typ} \vert_{h=h_c}  && \oppropto_{L \to +\infty} - L^{\psi} \ \ {\rm with } \ \ \psi \simeq 0.58  \nonumber \\
\ln h_L^{typ} \vert_{h>h_c} && \oppropto_{L \to +\infty} Cst 
\label{htyptriangle}
\end{eqnarray}

\subsection{ RG flow of the renormalized magnetization $\mu_L$ }

On Fig. \ref{figtriangleflowmuh} (b), we show the RG flow of the renormalized magnetization $\mu_L$
of surviving clusters as a function of $L$
\begin{eqnarray}
 \mu_L \vert_{h<h_c} && \oppropto_{L \to +\infty} L^D  \nonumber \\
 \mu_L  \vert_{h=h_c} && \oppropto_{L \to +\infty}  L^{d_f} \ \ {\rm with } \ \  d_f \simeq 1.08  \nonumber \\
 \mu_L \vert_{h>h_c} && \oppropto_{L \to +\infty} Cst 
\label{mutriangle}
\end{eqnarray}
 At criticality, the exponent $x$ of the intensive magnetization of Eq. \ref{mcriti}
is thus of order
\begin{eqnarray}
x=D-d_f \simeq 0.5
\label{xtriangle}
\end{eqnarray}

\subsection{ Critical exponents in the disordered phase $h>h_c$ }

\begin{figure}[htbp]
%\begin{figure}
 \includegraphics[height=6cm]{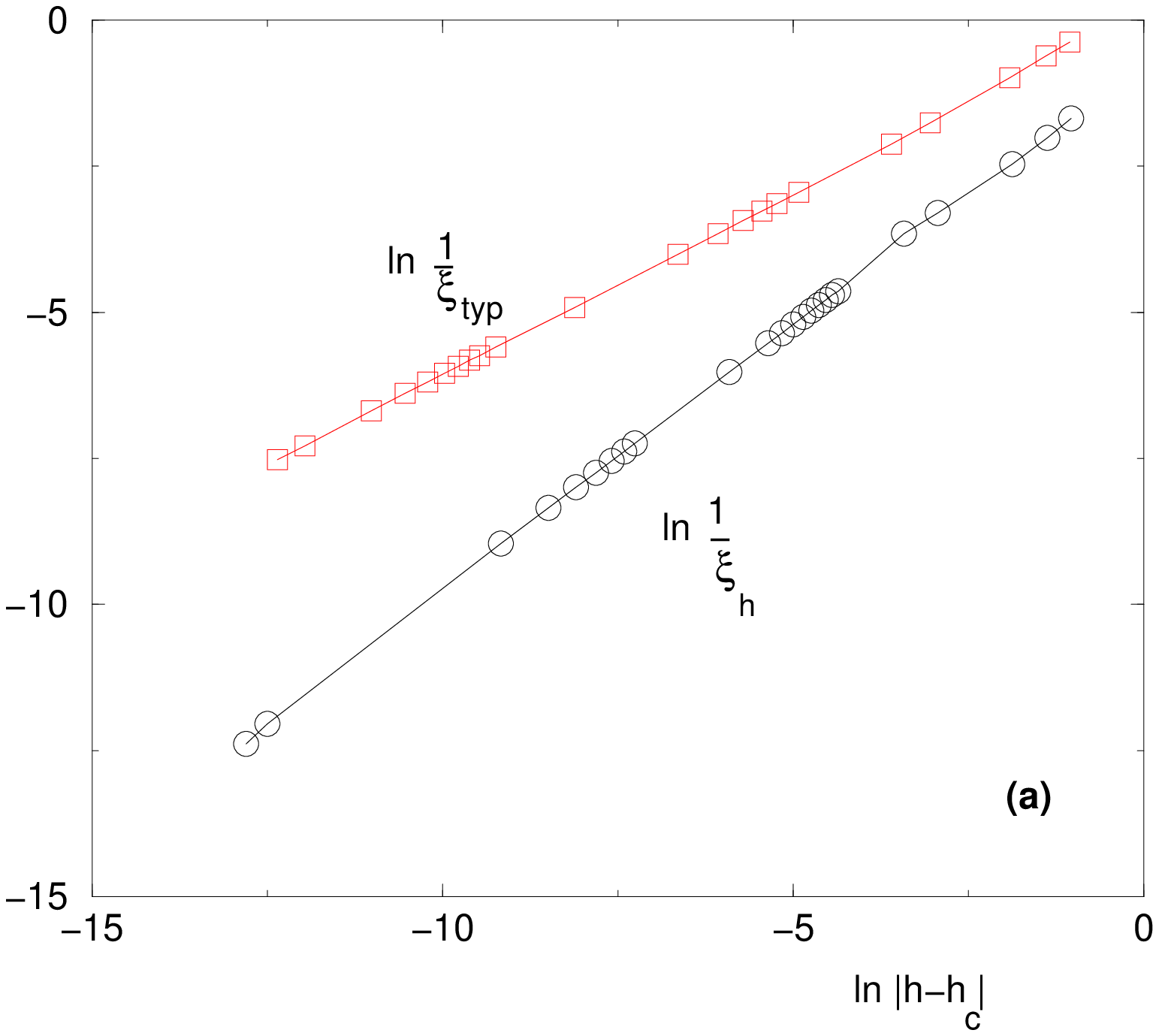}
\hspace{1cm}
\includegraphics[height=6cm]{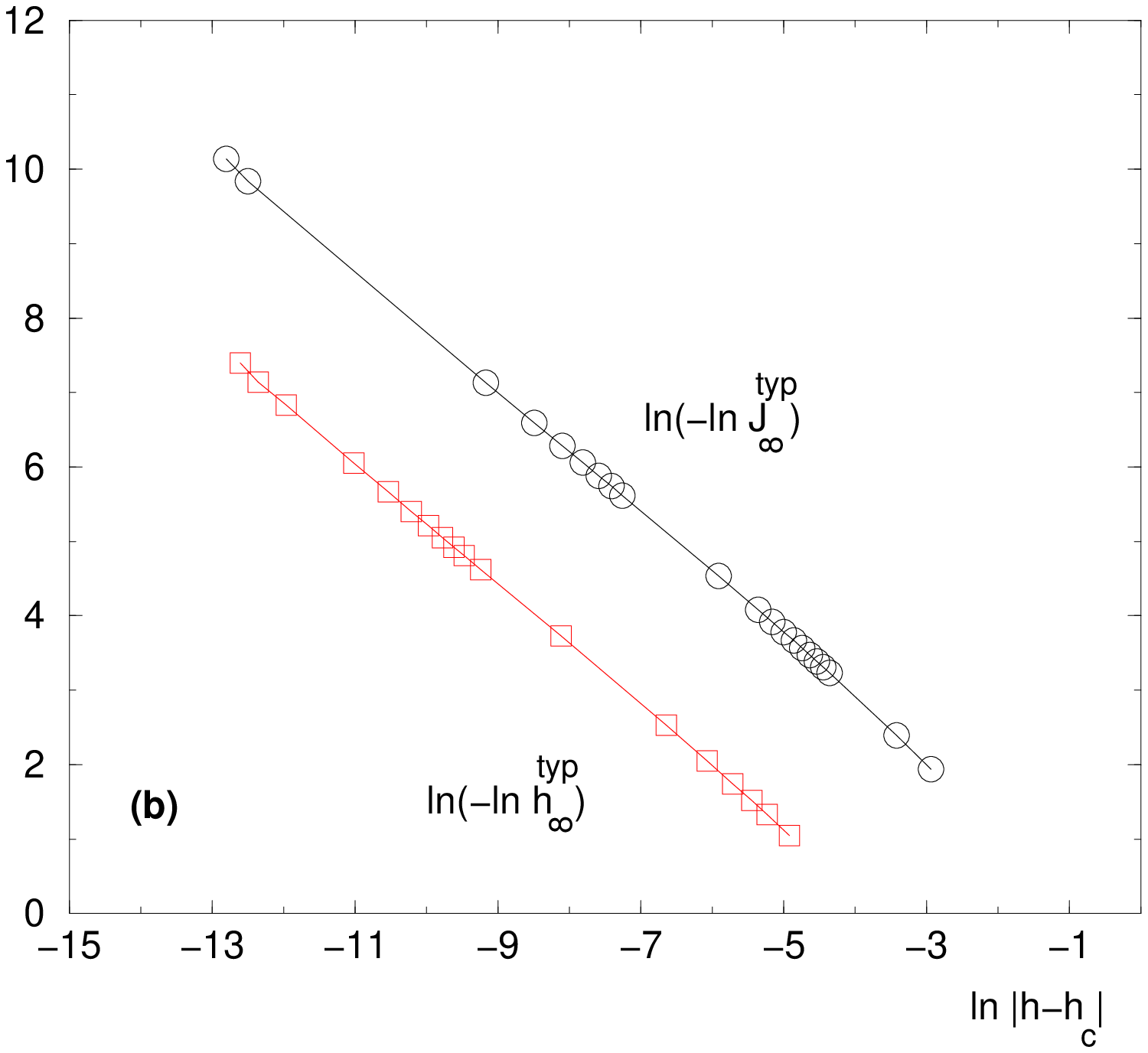}
\caption{ (Sierpinski ) 
(a)  Divergences of the typical correlation length $\xi_{typ}$ 
of the disordered phase and of the correlation length $\xi_h$ of the ordered phase :
we measure $\nu_{typ} \simeq 0.63 $ and  $\nu_h \simeq 0.91$
  (b) Essential singularities of the typical finite 
renormalized transverse field $h_{\infty}^{typ}$ in the disordered phase and 
of the typical finite 
renormalized coupling $J_{\infty}^{typ}$ in the ordered phase :
we measure $\kappa \simeq 0.82$ for both. }
\label{figtrianglexi}
\end{figure}

In the disordered phase, the exponential decay of the typical renormalized coupling
$J_L^{typ}=e^{\overline{\ln J_L }}$ defines the typical correlation length $\xi_{typ}$ (Eq. \ref{JLdisordertyp})
As shown on Fig. \ref{figtrianglexi} (a) , the divergence near criticality (Eq. \ref{nutypdes})
is governed by the exponent
\begin{eqnarray}
\nu_{typ} \simeq 0.63 
\label{nutyptriangle}
\end{eqnarray}

As shown on Fig. \ref{figtrianglexi} (b), the finite typical renormalized transverse field 
$h_{\infty}^{typ}$
displays the essential singularity of Eq. \ref{defkappa} with the value
\begin{eqnarray}
\kappa \simeq 0.82
\label{kappatriangle}
\end{eqnarray}

\subsection{ Critical exponents in the ordered phase $h<h_c$ }

\begin{figure}[htbp]
\includegraphics[height=6cm]{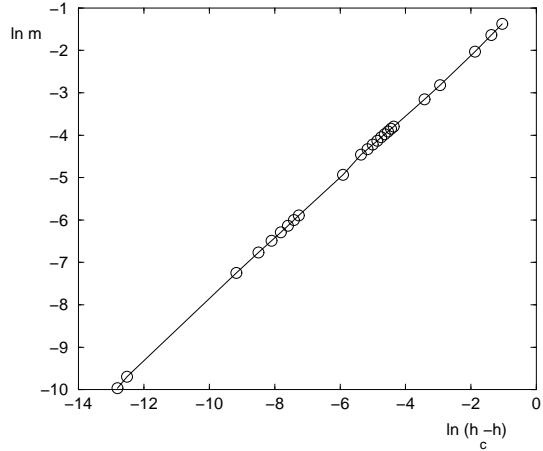}
\caption{ (Sierpinski )
 Singularity of the intensive magnetization $m$ in the ordered phase 
(Eq. \ref{defbeta}) :
we measure $\beta \simeq 0.73$
 }
\label{figtriangleorder}
\end{figure}

In the ordered phase,  
the exponential decay of the typical renormalized transverse field
$h_L^{typ}= e^{\overline{\ln h_L }}$ defines some characteristic length $\xi_h$ (Eq \ref{hLorder}).
As shown on Fig. \ref{figtrianglexi} (a), we find that $\xi_h$ diverges with the exponent
\begin{eqnarray}
\nu_h \simeq 0.91
\label{nuctriangle}
\end{eqnarray}

As shown on Fig. \ref{figtrianglexi} (a), 
 the finite typical renormalized coupling
$J_{\infty}^{typ}$ displays an essential singularity 
with the same exponent $\kappa \simeq 0.82$ as in Eq. \ref{kappatriangle}.

Our data concerning the intensive magnetization of Eq. \ref{defbeta}
are shown on Fig \ref{figtriangleorder} : we measure an exponent of order $\beta \simeq 0.73$.

Our various measures are thus compatible with a finite-size correlation exponent of order
\begin{eqnarray}
\nu_{FS} \simeq 1.45
\label{nufstriangle}
\end{eqnarray}

\section{ Numerical study for a hierarchical lattice having an underlying classical transition }

\label{sec_b2c8}

\begin{figure}[htbp]
\includegraphics[height=5cm]{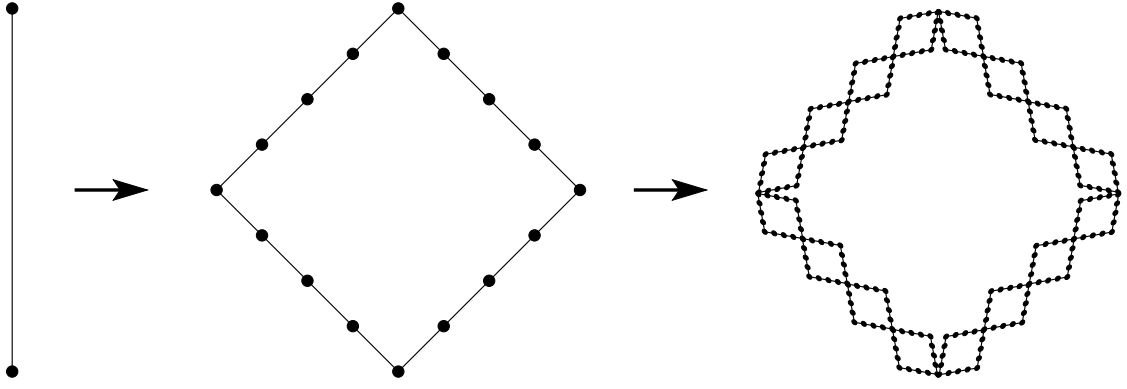}
\caption{ Hierarchical construction of the diamond lattice of
 branching ratio $k=2$ and rescaling factor $b=8$ :
we show the generation $n=0$ with a single bond, the generation $n=1$ 
containing  $kb=16$ bonds organized in $k=2$ branches of $b=8$ bonds in series,
and generation $n=2$ obtained by iteration.  }
 \label{figc8}
 \end{figure}

Fractal lattices present a classical ferromagnetic transition only if their ramification is infinite
\cite{gefen}. Exactly renormalizable lattices with a finite connectivity have a finite ramification
and do not present a classical ferromagnetic transition, as the Sierpinski gasket discussed in the
previous section. Exactly renormalizable lattices with a classical ferromagnetic transition
are thus hierarchical lattices presenting a growing connectivity, such as the lattice shown on Fig. \ref{figc8}
that we study in this section :
this lattice is constructed recursively
from a single link called here generation $n=0$.
At generation $n=1$, this single link has been replaced by $k=2$ branches, each branch
 containing $b=8$ bonds in series.
The generation $n=2$ is obtained by applying the same transformation
to each bond of the generation $n=1$.
At generation $n$, the length $L_n$ between the two extreme sites
$A$ and $B$ is $L_n=b^n$, whereas the total number of bonds grows as $N_n=(kb)^n = L_n^{D}$
so that the fractal dimension reads
\begin{eqnarray}
D= \frac{ \ln (kb)}{\ln b} = \frac{4}{3}
\label{dc8}
\end{eqnarray}

This type of hierarchical lattices has been much studied in relation with 
Migdal-Kadanoff block renormalizations \cite{MKRG} that can be considered in two ways, 
 either as approximate real space renormalization procedures on hypercubic lattices,
or as exact renormalization procedures on certain hierarchical lattices
\cite{berker,hierarchical}.
Various classical disordered models have been studied on these hiearchical lattice, in particular
 the diluted Ising model \cite{diluted}, 
the ferromagnetic random Potts model \cite{Kin_Dom,Der_Potts,andelman,diamondtails,diamondcriti,igloi_turban},
spin-glasses \cite{young,mckay,Gardnersg,bray_moore,nifle_hilhorst,diamondtails,diamondcriti}
and the directed polymer model
 \cite{Der_Gri,Coo_Der,Tim,roux,kardar,cao, tang,Muk_Bha,Bou_Sil,diamondtails,
diamondcriti}.

As in the previous sections, we have used the flat initial distribution of Eq. \ref{piJBox}
for the random couplings and a uniform transverse field $h$.
The results presented in this Section have been obtained
with a pool of size $M=10^7$. We find that the corresponding transition point satisfies
\begin{eqnarray}
0.556626170  < h_c^{pool}(M=10^7) <h=0.556626171
\label{hcpoolc8}
\end{eqnarray}
using $0 \leq n \leq 42$ generations
corresponding to lengths 
\begin{eqnarray}
1 \leq L=8^n \leq 8^{42}
\label{sizesc8}
\end{eqnarray}

\subsection{ RG flow of the typical renormalized coupling $J_L^{typ}$ }

\begin{figure}[htbp]
%\begin{figure}
 \includegraphics[height=6cm]{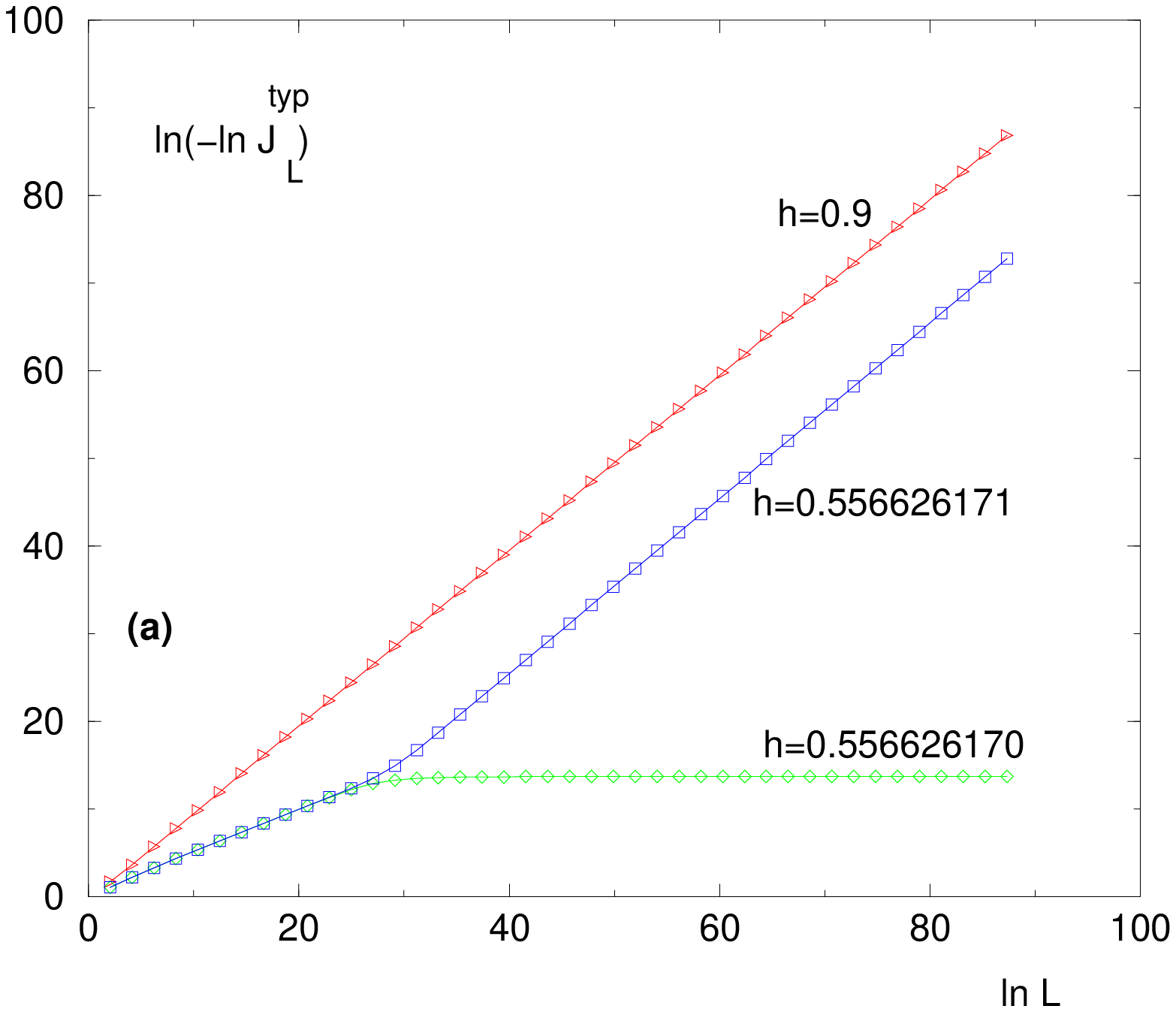}
\hspace{1cm}
 \includegraphics[height=6cm]{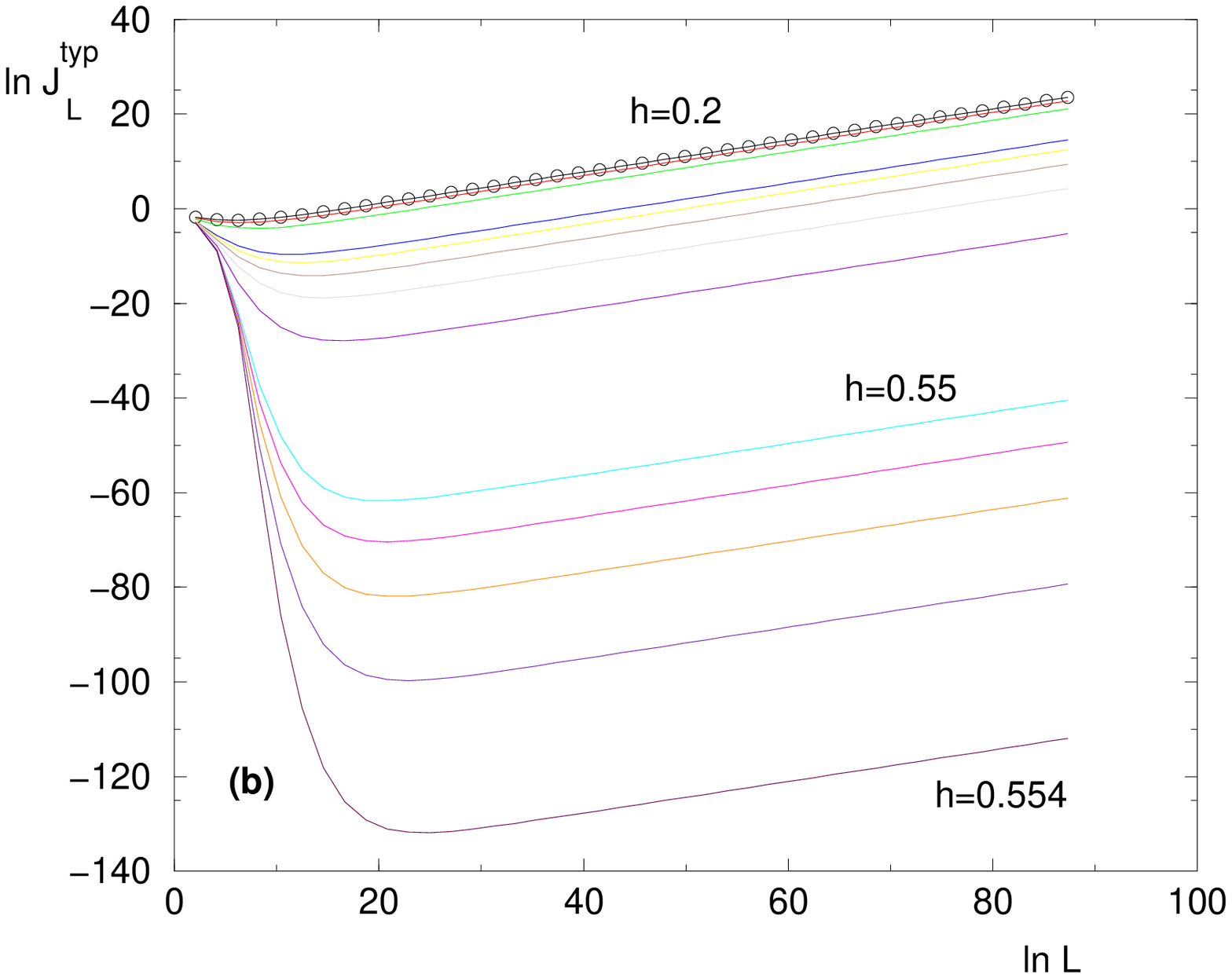}
\caption{ (Diamond $k=2$, $b=8$) 
(a)  RG flow of the logarithm of the typical renormalized coupling $(\ln J_L^{typ})$ in a log-log plot :
  the slope is $1$ in the disordered phase (e.g. $h=0.9$),
and the slope is $\psi \simeq 0.49 $ at criticality (before the bifurcation 
of the curves corresponding to $h=0.556626170$ and $h=0.556626171$ ).
(b)  RG flow in the ordered phase : the asymptotic growth corresponds to the slope
 $D_s=D-1=1/3$ (see Eq. \ref{jtypc8}).
 }
\label{figc8rgflowj}
\end{figure}

On Fig. \ref{figc8rgflowj}, we show our data concerning the RG flow of the typical coupling 
$J_L^{typ}=e^{\overline{\ln J_L }}$ at a function of $L$
\begin{eqnarray}
\ln J_L^{typ} \vert_{h<h_c} && \oppropto_{L \to +\infty} D_s \ln L   \ \ {\rm with } \ \ D_s=D-1=1/3 \nonumber \\
\ln J_L^{typ} \vert_{h=h_c}  && \oppropto_{L \to +\infty} - L^{\psi} \ \ {\rm with } \ \ \psi \simeq 0.49  \nonumber \\
\ln J_L^{typ} \vert_{h>h_c} && \oppropto_{L \to +\infty} -L 
\label{jtypc8}
\end{eqnarray}
The novelty with respect to the previous cases is thus the asymptotic growth 
$J_L^{typ} \propto L^{D_s}$ in the ordered phase, 
as expected when there exists an underlying classical transition (see the discussion 
around Eq. \ref{jijclassical}).

\subsection{ RG flow of the width $\Delta_L$ of the logarithms of the renormalized couplings }

\begin{figure}[htbp]
%\begin{figure}
 \includegraphics[height=6cm]{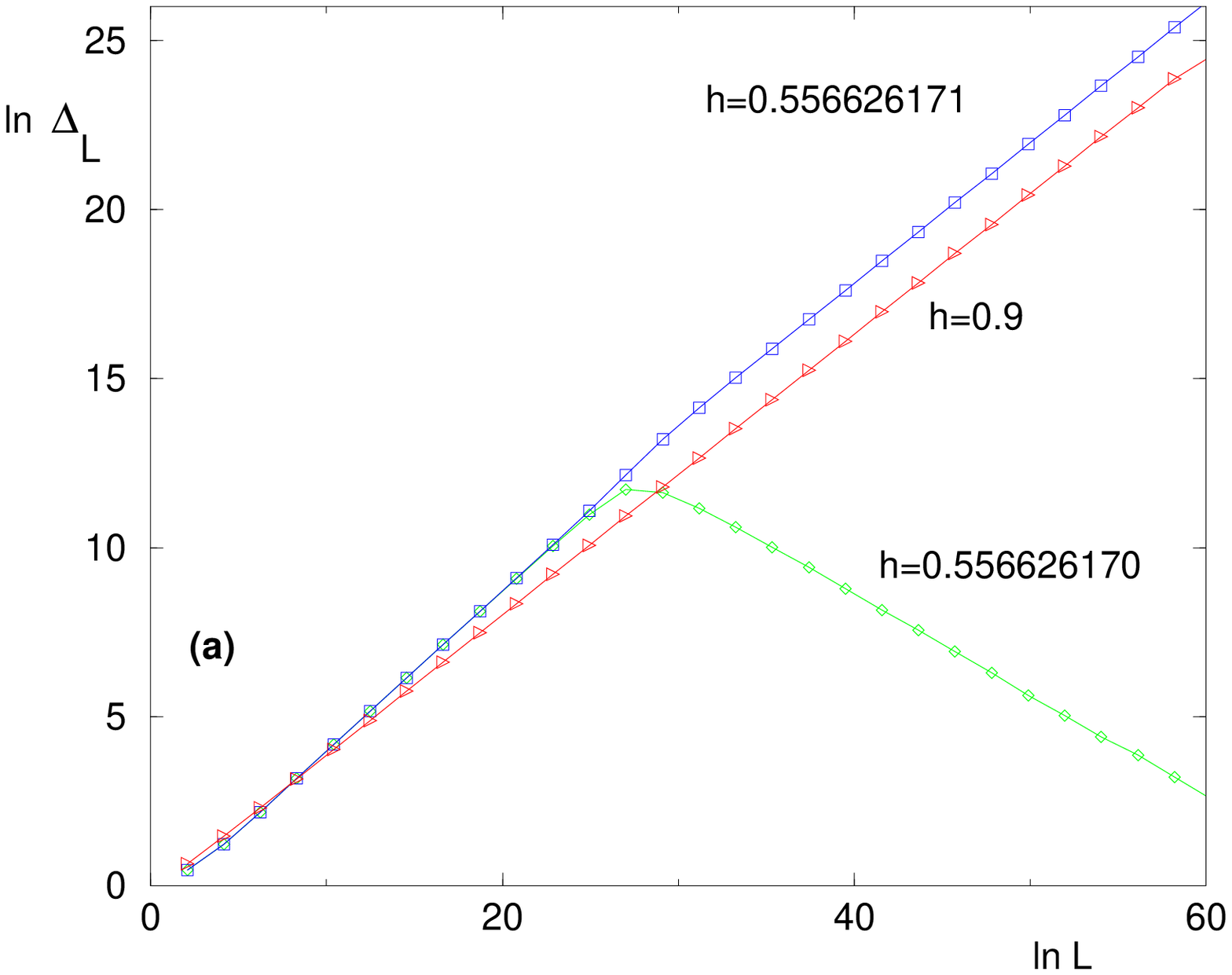}
\hspace{1cm}
 \includegraphics[height=6cm]{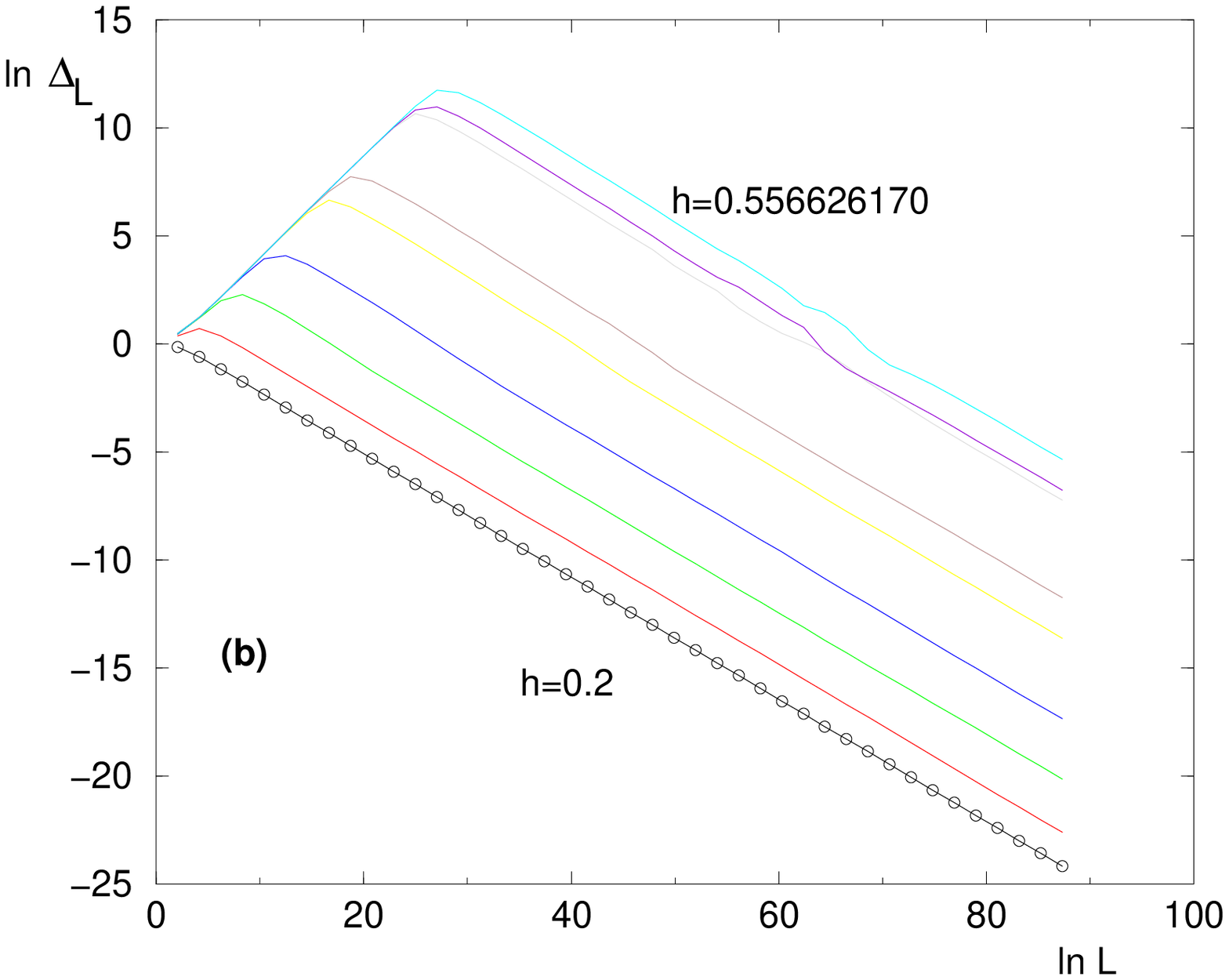}
\caption{ (Diamond $k=2$, $b=8$) 
(a)  RG flow of the width $\Delta_L$ of the distribution of the logarithm of the
renormalized couplings in a log-log plot : 
the slope is $\omega \simeq 0.41$ in the disordered phase (e.g. $h=0.9$),
and the slope is $\psi \simeq 0.49$ at criticality (before the bifurcation 
of the curves corresponding to $h=0.556626170$ and $h=0.556626171$
(b)  In the ordered phase, the width $\Delta_L$ decays asymptotically as $L^{\omega_s-D_s}$ with $\omega_s-D_s \simeq -0.29$ (see Eq. \ref{deltaLc8})
  }
\label{figc8rgflowdelta}
\end{figure}

On Fig. \ref{figc8rgflowdelta}, we show our data concerning the RG flow of the width $\Delta_L$
 of the distribution of the logarithms of the couplings
\begin{eqnarray}
\Delta_L  \vert_{h<h_c} && \oppropto_{L \to +\infty} L^{\omega_s-D_s}   \ \ {\rm with } \  \omega_s-D_s \simeq -0.29  \nonumber \\
\Delta_L  \vert_{h=h_c} && \oppropto_{L \to +\infty}  L^{\psi} \ \ {\rm with } \ \ \psi \simeq 0.48  \nonumber \\
\Delta_L  \vert_{h>h_c} && \oppropto_{L \to +\infty} L^{\omega} \ \ {\rm with } \ \ \omega \simeq 0.41
\label{deltaLc8}
\end{eqnarray}
Here the novelty with respect to the previous cases is again 
the ordered phase with the decay of the width $\Delta_L \propto L^{\omega_s-D_s}$
as expected when there exists an underlying classical transition (see the discussion 
around Eq. \ref{jijclassical}).

Another important result is the inequality $\psi>\omega$ implying a singular diverging
amplitude
$A$ for the width in the disordered phase (see the discussion around Eq. \ref{defampliA}).

\begin{figure}[htbp]
%\begin{figure}
 \includegraphics[height=6cm]{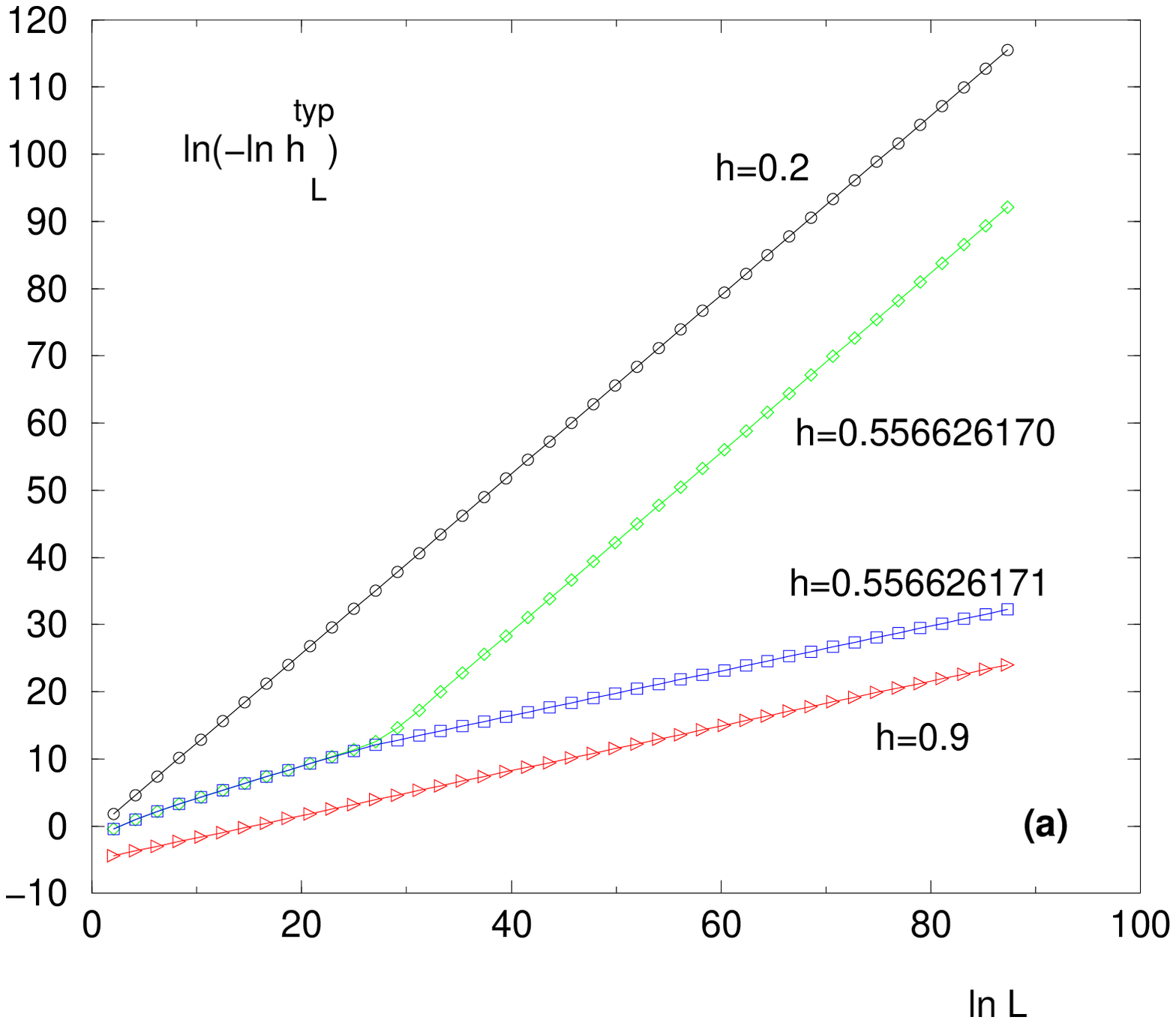}
\hspace{1cm}
 \includegraphics[height=6cm]{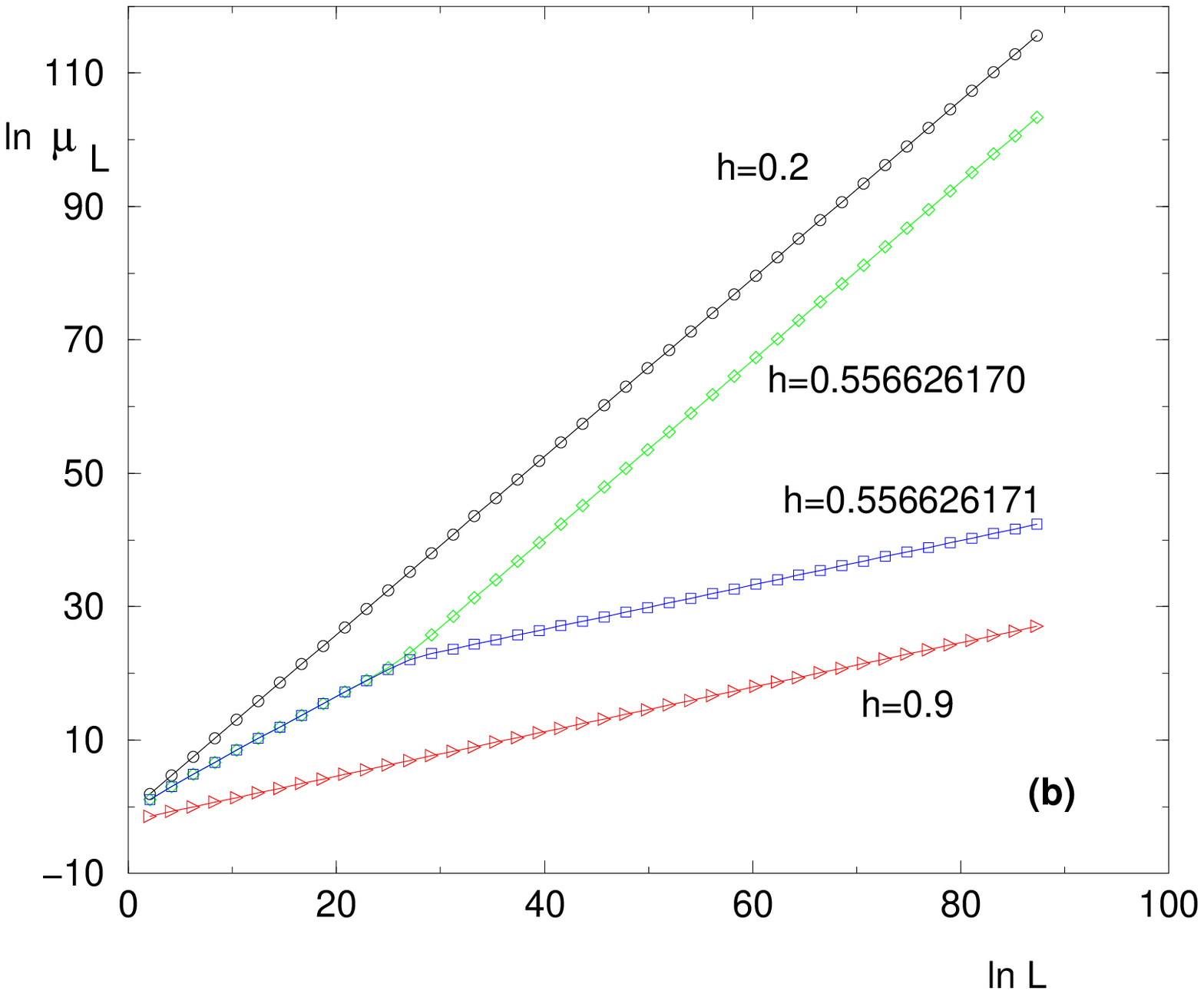}
\caption{ (Diamond $k=2$, $b=8$)
(a)  RG flow of the logarithm of the typical renormalized transverse field $(\ln h_L^{typ})$   in a log-log plot :
  the slope is $D=4/3$ in the ordered phase (e.g. $h=0.2$),
  the slope is $D_s=1/3$ in the disordered phase (e.g. $h=0.9$),
and the slope is $\psi \simeq 0.49 $ at criticality (before the bifurcation 
of the curves corresponding to $h=0.556626170$ and $h=0.556626171$ ).
(b)  RG flow of the renormalized magnetization $\mu_L$ in a log-log plot : 
the slope is $D=4/3$ in the ordered phase (e.g. $h=0.2$),
the slope is $D_s=1/3$ in the disordered phase (e.g. $h=0.9$),
and the slope is $d_f \simeq 0.85$ at criticality (before the bifurcation 
of the curves corresponding to $h=0.556626170$ and $h=0.556626171$).
  }
\label{figc8rgflowmuh}
\end{figure}

\subsection{ RG flow of the typical renormalized transverse field $h_L^{typ}$}

On Fig. \ref{figc8rgflowmuh} (a), we show the RG flow of $h_L^{typ} $
as a function of the RG scale $L$ 
\begin{eqnarray}
\ln h_L^{typ}  \vert_{h<h_c} && \oppropto_{L \to +\infty} -L^D  \nonumber \\
\ln h_L^{typ}  \vert_{h=h_c} && \oppropto_{L \to +\infty} - L^{\psi} \ \ {\rm with } \ \ \psi \simeq 0.49  \nonumber \\
\ln h_L^{typ} \vert_{h>h_c} && \oppropto_{L \to +\infty} -L^{D_s}  \ \ {\rm with } \ \ D_s = D-1 =1/3
\label{htypc8}
\end{eqnarray}
The novelty with respect to the previous cases is 
the disordered phase that involves the dimension $D_s$ as a consequence of the growing connectivity
of the diamond hierarchical lattice (see the discussion in \ref{sec_ds}).

\subsection{ RG flow of the renormalized magnetization $\mu_L$ of surviving clusters}

On Fig. \ref{figc8rgflowmuh} (b), we show the RG flow of the renormalized magnetization $\mu_L$ of surviving clusters
as a function of the RG scale $ L$ 
\begin{eqnarray}
 \mu_L \vert_{h<h_c} && \oppropto_{L \to +\infty} L^D  \nonumber \\
 \mu_L  \vert_{h=h_c}  && \oppropto_{L \to +\infty}  L^{d_f} \ \ {\rm with } \ \  d_f \simeq 0.85  \nonumber \\
 \mu_L \vert_{h>h_c} && \oppropto_{L \to +\infty}  L^{D_s}  \ \ {\rm with } \ \ D_s = D-1 =1/3
\label{muc8}
\end{eqnarray}
 At criticality, the exponent $x$ of the intensive magnetization of Eq. \ref{mcriti}
is thus of order
\begin{eqnarray}
x=D-d_f \simeq 0.48
\label{xc8}
\end{eqnarray}

\subsection{ Critical exponents in the disordered phase $h>h_c$ }

\begin{figure}[htbp]
 \includegraphics[height=6cm]{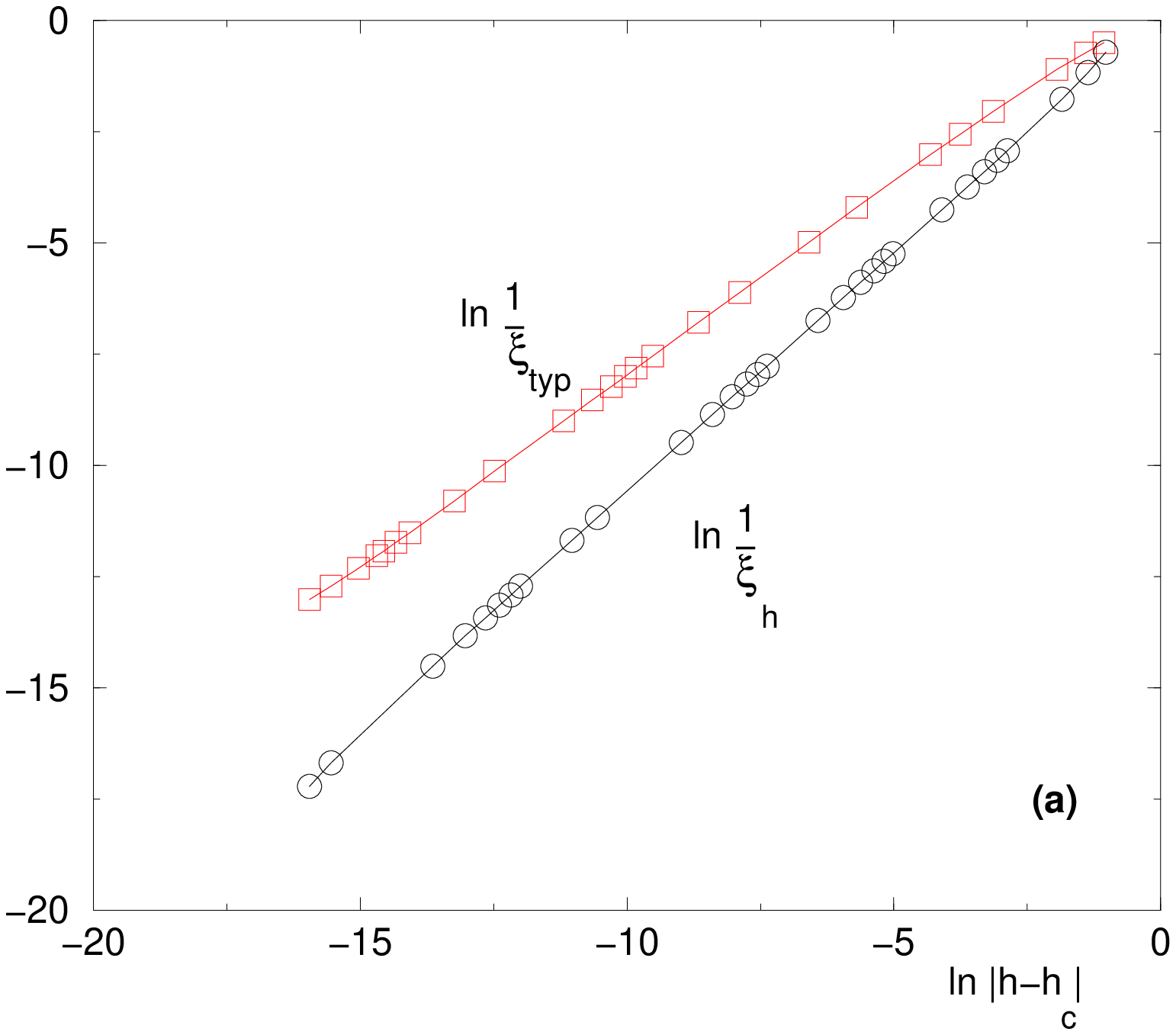}
\hspace{1cm}
 \includegraphics[height=6cm]{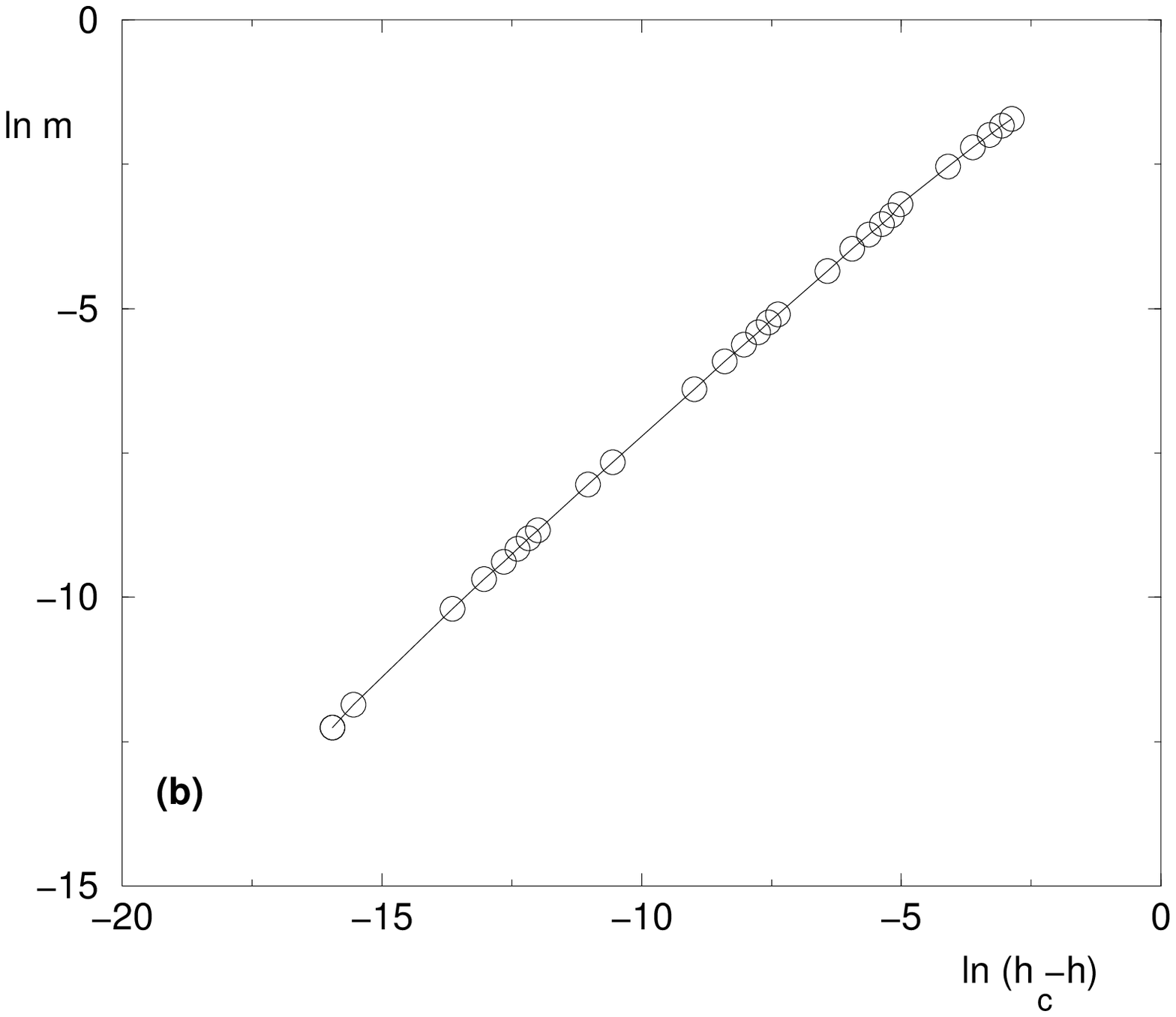}
\caption{ (Diamond $k=2$, $b=8$) Critical exponents 
(a) Divergences of the typical correlation length $\xi_{typ}$ 
of the disordered phase and of the correlation length $\xi_h$ of the ordered phase :
we measure $\nu_{typ} \simeq 0.86$ and $\nu_h \simeq 1.08 $
(b)  Singularity of the intensive magnetization $m$ in the ordered phase 
(Eq. \ref{defbeta}) :
we measure $\beta \simeq 0.81$. }
\label{figc8exponents}
\end{figure}

In the disordered phase, the exponential decay of the typical renormalized coupling
$J_L^{typ}=e^{\overline{\ln J_L}}$ defines the typical correlation length $\xi_{typ}$
As shown on Fig. \ref{figc8exponents} (a), the divergence near criticality
is governed by an exponent of order
\begin{eqnarray}
\nu_{typ} =0.86
\label{nutypc8}
\end{eqnarray}

\subsection{ Critical exponents in the ordered phase $h<h_c$ }

In the ordered phase,  
the exponential decay of the typical renormalized transverse field
$h_L^{typ}= e^{\overline{\ln h_L }}$ defines the typical cluster linear length $\xi_h$
\begin{eqnarray}
\ln  h_L^{typ}\equiv \overline{\ln h_L } \opsimeq_{L \to +\infty} - \left( \frac{L}{\xi_h} \right)^D
\label{xityporderc8}
\end{eqnarray}
where $\xi_h$ diverges with the exponent $\nu_h \simeq 1.08$ (see Fig \ref{figc8exponents} (a))

On Fig. \ref{figc8exponents} (b), we show our data concerning the intensive magnetization  
 : we measure an exponent of order $\beta \simeq 0.81$.

Our various measures are thus compatible with a finite-size correlation exponent of order 
\begin{eqnarray}
\nu_{FS} \simeq 1.71
\label{nufsc8}
\end{eqnarray}

\subsection{ Discussion :  effect of the growing connectivity }

\label{sec_ds}

In this section, we have considered a hierarchical fractal lattice,
where the end points have a growing connectivity as $k^n$, when the length grows as $L_n=b^n$.
This property has for consequence that in the disordered phase,
the magnetization $\mu_L$ and the logarithms of the transverse fields
do not remain finite as they do when they represent sites of finite connectivity,
 but scale as
the growing connectivity as $k^n=L_n^{D_s}$ where
\begin{eqnarray}
D_s= \frac{\ln k}{\ln b} = D-1
\label{ds}
\end{eqnarray}

By consistency in the disordered phase, only sites should be decimated asymptotically,
i.e. the logarithms of the renormalized transverse fields scaling as
\begin{eqnarray}
\ln h_L \oppropto_{L \to +\infty} - L^{D_s}
\label{lnhLds}
\end{eqnarray}
should remain bigger than the logarithm of the renormalized couplings scaling as
in Eq. \ref{JLdisorder} : the RG procedure is thus consistent only if $D_S<1$.
In addition, if one wishes that the spurious behavior of Eq. \ref{lnhLds}
does not affect the critical point either, one should have the stronger inequality
\begin{eqnarray}
D_s < \psi
\label{dspsi}
\end{eqnarray}
This condition is satisfied in the case we have studied numerically with $D_s \simeq 0.33 < 0.49 \simeq \psi$ (and this is why we have chosen the value $b=8$ instead of smaller values).
But it is clear that this condition will not be satisfied by all diamond hierarchical lattices with arbitrary values of the parameters $(k,b)$ of Eq. \ref{dc8}.

\section{ Conclusion}

\label{sec_conclusion}

 In this paper, we have proposed to include Strong Disorder
RG ideas within the more traditional fixed cell-size real space RG framework.
We have first considered the one-dimensional chain as a test for this fixed 
cell-size procedure.
Our conclusion is that all exactly known critical exponents are reproduced correctly, except for the magnetic exponent $\beta$, because it is related to more subtle persistence properties of the full RG flow. 
We have then applied numerically
this fixed cell-size RG procedure to two types of renormalizable fractal lattices :

(i) the Sierpinski gasket of fractal dimension $D=\ln 3/\ln 2$, where there is no underlying classical ferromagnetic transition, so that the RG flow in the ordered phase is similar to what happens in $d=1$.

(ii) a hierarchical diamond lattice of fractal dimension $D=4/3$, where there is an underlying classical ferromagnetic transition, so that the RG flow in the ordered phase is similar to what happens on hypercubic lattices of dimension $d>1$. 

In both cases, we have found that the transition is governed by an Infinite Disorder Fixed Point. Besides the measure of the activated exponent $\psi$ at criticality, we have analyzed the RG flow of various observables in the disordered phase and in the ordered phase, in order to extract the
'typical' correlation length exponents of these two phases (respectively $\nu_{typ}$ of Eq. \ref{JLdisordertyp}
and $\nu_h$ of Eq. \ref{hLorder}) which are different from the finite-size correlation exponent $\nu_{FS}$
that governs all finite-size properties in the critical region (Eq. \ref{fsslnjl}). In the disordered phase,
we have also measured the fluctuation exponent $\omega$ (Eq. \ref{JLdisorder}) which is expected to
 coincide with the droplet exponent of Directed Polymer living on the same lattice \cite{us_transverseDP}.
Whereas the two exponents $\psi$ and $\omega$ are known to coincide in dimension $d=1$, we have found here
in section \ref{sec_b2c8} that it is possible to have $\omega<\psi$ (and accordingly a diverging amplitude
$A(h)$ in Eq. \ref{JLdisorder}). Moreover the measured value $\psi \simeq 0.49$ is close to the values
measured for hypercubic lattices in dimensions $d=2,3,4$  \cite{fisherreview,motrunich,lin,karevski,lin07,yu,kovacsstrip,kovacs2d,kovacs3d,kovacsentropy,kovacsreview}.

Besides the random transverse field Ising model that we have discussed, we hope that the idea to include 
Strong Disorder RG principles within a fixed cell-size real space renormalization
will be useful to study other types of disordered models on fractal lattices,
like for instance random walks with disorder on various fractal lattices 
that have been studied recently in \cite{rwfractal}.

\appendix

\section{ Reminder on the full Strong Disorder RG procedure in energy}

\label{sec_full}

In this section, we recall the standard Strong Disorder Renormalization for 
 the Random Transverse Field Ising Model of Eq. \ref{hdes}, to compare with the 
fixed cell-size framework introduced in the text.

\subsection{ Reminder on Strong Disorder RG rules on arbitrary lattices }

\label{SDRGfullrules}

For the model of Eq. \ref{hdes},
the Strong Disorder RG rules are formulated on arbitrary lattices as follows \cite{fisherreview,motrunich} :

(0) Find the maximal value among the transverse fields $h_i$
and the ferromagnetic couplings $J_{jk}$
\begin{eqnarray}
\Omega= {\rm max } \left[h_i,J_{jk} \right]
\label{omega}
\end{eqnarray}

i) If $\Omega= h_{i}$, then the site $i$ is decimated and disappears,
while all couples $(j,k)$ of neighbors of $i$ are now linked via
the renormalized ferromagnetic coupling
\begin{eqnarray}
J_{jk}^{new} = J_{jk} + \frac{ J_{ji} J_{ik} }{h_{i}}
\label{jjknew}
\end{eqnarray}

ii) If $\Omega= J_{ij}$, then the site $j$ is merged with the site $i$.
The new renormalized site $i$ has a reduced renormalized transverse field
\begin{eqnarray}
h_{i}^{new}= h_i r_i \ \ {\rm with } \ \ r_i= \frac{h_j}{ J_{ij}  }
\label{hinew}
\end{eqnarray}
and a bigger magnetic moment
\begin{eqnarray}
 \mu_{i}^{new}=\mu_{i}+\mu_j
\label{minew}
\end{eqnarray}
This renormalized cluster is connected to other sites via the renormalized couplings
\begin{eqnarray}
J_{ik}^{new} =  J_{ik}+J_{jk}
\label{jiknew}
\end{eqnarray}

(iii) return to (0).

\subsection{ RG flow of probability distributions }

The above RG rules determine the RG flow for the joint probability distribution
at scale 
\begin{eqnarray}
\Omega=e^{-\Gamma}
\label{defgamma}
\end{eqnarray}
 of the renormalized clusters characterized by their magnetic moments $\mu_i$,
their renormalized fields 
\begin{eqnarray}
h_i=\Omega e^{-\beta_i} = e^{-(\Gamma+\beta_i)}
\label{defbetai}
\end{eqnarray}
 and the renormalized couplings between them 
\begin{eqnarray}
J_{ij}=\Omega e^{-\zeta_{ij}} = e^{-(\Gamma+\zeta_{ij})}
\label{defzetaij}
\end{eqnarray}
where $\beta_i$ and $\zeta_{ij}$ are positive random variables.
In the following, we recall \cite{motrunich,fisherreview} the properties of the
probability distributions $R_{\Gamma}(\beta_i )$ and $P_{\Gamma}(\zeta_{ij})$
(normalized per surviving cluster)
and of the density $n_{\Gamma}$ of surviving clusters per unit volume that defined the characteristic
length-scale $l_{\Gamma}$ via
\begin{eqnarray}
n_{\Gamma} \sim \frac{1}{l_{\Gamma}^d }
\label{ngamma}
\end{eqnarray}

\subsection{ Disordered phase }

\label{sec_disorder}

In the disordered phase, renormalized clusters remain 'finite', i.e.
asymptotically at large RG scale $\Gamma$, only transverse fields are decimated
via the rule (i) of section \ref{SDRGfullrules}, while 
the rule (ii) does not occur anymore. 
This means that the variable $\beta_i$ of Eq. \ref{defbetai} remains finite,
i.e.  the probability distribution $R_{\Gamma}(\beta_i )$ 
converges towards a finite probability distribution $R_{\infty}(\beta_i )$
without any rescaling for the variable $\beta_i$.
The value $R_{\infty}(0 )$ at the origin $\beta=0$ will then govern
the decay of the clusters density $n_{\Gamma}$ 
\begin{eqnarray}
\partial_{\Gamma} n_{\Gamma} = - R_{\infty}(0 ) n_{\Gamma}
\label{dngammadisordered}
\end{eqnarray}
leading to the exponential decay
\begin{eqnarray}
 n_{\Gamma} \propto e^{ - \Gamma R_{\infty}(0 ) }
\label{ngammadisordered}
\end{eqnarray}
or equivalently to the exponential growth of the length-scale $l_{\Gamma}$ of Eq. \ref{ngamma}
representing the typical distance between surviving clusters
\begin{eqnarray}
 l_{\Gamma} \propto e^{  \Gamma \frac{R_{\infty}(0 )}{d} }
\label{lgammadisordered}
\end{eqnarray}
The relation between the energy scale $\Omega=e^{-\Gamma}$ and the length scale $l_{\Gamma}$
thus corresponds to the power-law
\begin{eqnarray}
\Omega=e^{-\Gamma} \propto l_{\Gamma}^{-z}  
\label{zdefdisordered}
\end{eqnarray}
with the continuously variable dynamical exponent
\begin{eqnarray}
z = \frac{d}{R_{\infty}(0 )}
\label{zdisordered}
\end{eqnarray}

\subsection{ Critical point }

\label{sec_criti}

At the critical point, both transverse fields and couplings continue to be decimated at large scale $\Gamma$,
and there is no finite characteristic scale :
as a consequence, the only scale available for the variables $\beta_i$ of Eq. \ref{defbetai}
and $\zeta_{ij}$ of Eq. \ref{defzetaij} is the RG scale $\Gamma$ itself.
More precisely, the probability distributions $R_{\Gamma}(\beta_i )$ 
and $P_{\Gamma}(\zeta_{ij})$ (normalized per surviving cluster) follow the scaling form
\begin{eqnarray}
R_{\Gamma}(\beta_i ) && \simeq \frac{1}{\Gamma} {\cal R} \left( \frac{\beta_i}{\Gamma} \right) \nonumber \\
P_{\Gamma}(\zeta_{ij}) && \simeq \frac{1}{\Gamma} {\cal P} \left( \frac{\zeta_{ij}}{\Gamma} \right)
\label{prescalcriti}
\end{eqnarray}
where the rescaled probability distributions ${\cal R}$ and ${\cal P}$ do not depend on $\Gamma$.

The values ${\cal R}(0) $ and ${\cal P}(0) $ at the origin will then govern
the decay of the clusters density $n_{\Gamma}$ 
\begin{eqnarray}
\partial_{\Gamma} n_{\Gamma} = - \left( \frac{{\cal R}(0)}{\Gamma} + \frac{{\cal P}(0)}{\Gamma} \right)  n_{\Gamma}
\label{dngammacriti}
\end{eqnarray}
leading to the power-law
\begin{eqnarray}
 n_{\Gamma} \propto \frac{1}{\Gamma^{{\cal R}(0)+{\cal P}(0)}}
\label{ngammacriti}
\end{eqnarray}
or equivalently to the power-law growth of the length-scale $l_{\Gamma}$ of Eq. \ref{ngamma}
\begin{eqnarray}
 l_{\Gamma} \propto \Gamma^{\frac{{\cal R}(0)+{\cal P}(0)}{d}}
\label{lgammacriti}
\end{eqnarray}
The relation between the energy scale $\Omega=e^{-\Gamma}$ and the length scale $l_{\Gamma}$
thus corresponds to the activated form
\begin{eqnarray}
\Omega=e^{-\Gamma} \propto e^{-  l_{\Gamma}^{\psi}} 
\label{defpsicriti}
\end{eqnarray}
where the activated exponent $\psi$ reads
\begin{eqnarray}
\psi = \frac{d}{{\cal R}(0)+{\cal P}(0)}
\label{psicriti}
\end{eqnarray}

Finally, the magnetic moment of surviving clusters is expected to grow as a power-law of $\Gamma$,
and equivalently as some power-law of $l_{\Gamma}$ (Eqs \ref{lgammacriti} and \ref{psicriti})
\begin{eqnarray}
\mu \propto \Gamma^{\Phi} \propto l_{\Gamma}^{d_f} \ \ {\rm with } \ \ d_f=\Phi \psi
\label{mucriti}
\end{eqnarray}
where $d_f$ represents the fractal dimension of surviving critical clusters.

\subsection{ Ordered phase }

\label{sec_order}

In the ordered phase, renormalized clusters are expected to become 'extensive',
 i.e. asymptotically at large RG scale $\Gamma$, only couplings are decimated
via the rule (ii) of section \ref{SDRGfullrules}, while 
the rule (i) does not occur anymore. 
For the asymptotic behavior of the renormalized couplings $J_{ij}$ between surviving clusters,
one has to distinguish two possible cases, depending on the existence or non-existence
of an underlying classical ferromagnetic phase.

\subsubsection{ Ordered phase when there is no underlying classical ferromagnetic phase (as in $d=1$) }

\label{sec_orderwithoutclassical}

 When there is no underlying classical ferromagnetic phase (as in the one-dimensional chain),
 the variables $\zeta_{ij}$  of Eq. \ref{defzetaij} remain finite,
i.e.  the probability distribution $P_{\Gamma}(\zeta_{ij} )$ (normalized per surviving clusters)
converges towards a finite probability distribution $P_{\infty}(\zeta_{ij})$
without any rescaling for the variable $\zeta_{ij}$.
The value $P_{\infty}(0 )$ at the origin $\zeta=0$ will then govern
the decay of the clusters density $n_{\Gamma}$ 
\begin{eqnarray}
\partial_{\Gamma} n_{\Gamma} = - P_{\infty}(0 ) n_{\Gamma}
\label{dngammaorderedcase1}
\end{eqnarray}
leading to the exponential decay
\begin{eqnarray}
 n_{\Gamma} \propto e^{ - \Gamma P_{\infty}(0 ) }
\label{ngammaorderedcase1}
\end{eqnarray}
or equivalently to the exponential growth of the length-scale $l_{\Gamma}$ of Eq. \ref{ngamma}
representing the typical linear size of surviving clusters
\begin{eqnarray}
 l_{\Gamma} \propto e^{  \Gamma \frac{P_{\infty}(0 )}{d} }
\label{lgammaorderedcase1}
\end{eqnarray}
The relation between the energy scale $\Omega=e^{-\Gamma}$ and the length scale $l_{\Gamma}$
thus corresponds to the power-law of Eq. \ref{zdefdisordered}
with the continuously variable dynamical exponent
\begin{eqnarray}
z = \frac{d}{P_{\infty}(0 )}
\label{zorderedcase1}
\end{eqnarray}
i.e. the ordered phase is thus rather 'symmetric' to the disordered phase (see Eq \ref{zdisordered}).
Note that in dimension $d=1$, this symmetry is
a consequence of the duality between couplings and transverse fields \cite{danieltransverse}.

\subsubsection{ Ordered phase when there is an underlying classical ferromagnetic phase (as in $d>1$) }

 When there exists an underlying classical ferromagnetic phase
 (as on the hypercubic lattices in dimension $d>1$),
 the ordered phase is completely different
from the properties just described, because an infinite percolation cluster
appears at some finite RG scale $\Gamma_{perco}$ :
we refer to Ref \cite{motrunich} for more details.

\end{document}